\documentclass{JAC2003}
\usepackage{graphicx}
\usepackage{booktabs}

\usepackage[fleqn]{amsmath}
\usepackage{epsfig}
\begin{document}
\title{THE SYNCHROTRON MOTION SIMULATOR FOR 
ADIABATIC CAPTURE STUDY IN THE TLS BOOSTER}

\author{Cheng-Chin Chiang\thanks{chengchin.chiang@gmail.com}, Taipei, Taiwan}

\maketitle

\begin{abstract}
The synchrotron motion simulator is invented to simulate the 
longitudinal motion for particles under RF (Radio Frequency) voltage 
field in a ring accelerator. 
It is especially used to study the efficiency of adiabatic capture 
for a booster ring. 
The purpose of adiabatic capture is to optimize RF voltage settings 
during the ramping of beam energy and obtain the greatest efficiency 
of particle capture.
In this paper we study the longitudinal (synchrotron) motion for particles 
in the TLS (Taiwan Light Source) booster ~\cite{ref-TLS-B-1, ref-TLS-B-2}.
We compare the properties of TLS booster as a proton or electron 
accelerator, using the same ramping scenario of beam energy, 
and optimize the RF voltage settings to have the best capture efficiency.
\end{abstract}

\section{INTRODUCTION}
The TLS booster is a combined function FODO lattice with 
twelve periods, the circumference is 72 m.
Its original designed is to boost electrons energy from 
50 (MeV) to 1.5 (GeV).
It has been operated for almost twenty years.
The new accelerator TPS (Taiwan Photon Source) is planned to 
replace the TLS.
If the TLS storage ring is decided to decommission in future, 
the TLS booster can be considered to transfer to 
a proton accelerator for nuclear or medical researches.

The capture efficiency for a proton booster is required to be 
high enough to minimize the proton losses. 
Since the loss of protons would cause radiation contaminations 
and endanger the human body or environment. 
Given the ranges of RF operation and ramping scenario of the beam energy, 
we can search for the best settings for adiabatic capture by 
synchrotron motion simulator.
In the case of TLS booster, we plan to accelerate protons 
with the kinetic energy from 7 (MeV) to 300 (MeV) in the time period 50 (ms).
The operation range of RF voltage is from 0 to 15 (kV), and the 
harmonic number is 2. 

\section{EVOLUTION OF SYNCHROTRON PHASE-SPACE ELLIPSE}
The RF cavity is operated in a resonance condition to provide 
accelerating voltage, i.e. longitudinal electric field, to particles. 
For simplicity, we do not consider the effects of synchro-betatron coupling.
The synchrotron equations of motion can be derived from Hamiltonian.
For the beam acceleration, it is suitable to choose the phase-space 
mapping equation in the coordinates $(\phi,\Delta E)$, 
where $\phi$ is the phase of RF voltage and $\Delta E$ is the change of 
a particle energy per revolution.
Let $n$ is the turn number for a particle in a ring accelerator. 
The evolution equations of synchrotron motion in phase-space 
$(\phi,\Delta E)$ are derived from~\cite{ref-SYLee}: 
\begin{equation}
\label{eq:PhaseSpace}
\begin{split}
 & \Delta E_{n+1} = \Delta E_n + eV(\sin\phi_n - \sin\phi_s) 
                    - C_{\gamma}\frac{E^4}{\rho}, \\
 & \phi_{n+1} = \phi_n + \frac{2\pi h \eta}{\beta^2 E}\Delta E_{n+1}, \\
 & \mathrm{where\ } 
     \begin{cases}
        eV\sin\phi_s = E(t_{n+1}) - E(t_{n}), \\
        \gamma = \frac{E}{m_0c^2}, \\
        \beta = \sqrt{1-\frac{1}{\gamma^2}}, \\
        v = \beta c, \\
        \eta = \alpha_c - \frac{1}{\gamma^2}, \\
        \frac{\Delta E}{\beta^2 E} = \frac{\Delta P}{P}.
     \end{cases}
\end{split}
\end{equation}
$e$ is particle charge, 
$V$ is RF voltage, 
$E$ is the total energy of a particle, 
$\rho$ is the local radius of curvature for a bending magnet, 
$h$ is the harmonic number for RF, 
$\phi_s$ is the phase factor for RF, 
$m_0$ is the stationary mass of a particle, 
$c$ is the light speed, 
$v$ is particle velocity, 
$\alpha_c$ is the momentum compaction factor for a ring accelerator and 
$P$ is the momentum of a particle.
The radiation power coefficient $C_{\gamma}$ is deduced from 
Larmor's theorem, which dependents on different kind of particles: 
\begin{equation}
\label{eq:C_gamma}
\begin{split}
  C_{\gamma} & = \frac{4\pi}{3}\frac{r_0}{\left(m_0c^2\right)^3} \\
             & = \begin{cases}
                 8.846 \times 10^{-5} \rm{\ m/(GeV)^3 \ for \ electrons,} \\
                 7.783 \times 10^{-18} \rm{\ m/(GeV)^3 \ for \ protons,}
                 \end{cases}
\end{split}
\end{equation}
where $r_0=e^2/4\pi \epsilon_0 m_0 c^2$ is the classical radius for a particle.

Note that $V$ and $E$ can be as functions of time during the ramping 
of beam energy.
The RF voltage $V$ as a function of time in a ramping cycle 
is modeled according to~\cite{ref-Kang}:  
\begin{equation}
\begin{split}
\label{eq:V_RF}
  V(t) =
    \begin{cases}
         \left[ 3 \left( \frac{t}{T_{\nu}} \right)^2 
                - 2 \left( \frac{t}{T_{\nu}} \right)^3
         \right] \left(V_f-V_i\right) + V_i \\
    \mathrm{\ \ \ \ \ \ \ \ \ \ \ \ \ \ \ \ \ \ \ \ \ \ \
            \ \ \ \ \ \ \ \ \ \ \ \ \ \ \ \ for \ \ } 
    0\le t \le T_{\nu}, \\
    V_f \mathrm{\ \ for \ \ } t > T_{\nu}, 
    \end{cases}
\end{split}
\end{equation}
where $T_{\nu}$ the adiabatic capture time, 
$V_i$ and $V_f$ are initial and final RF voltages, respectively. 
The setting of $T_{\nu}$ would affect the efficiency for adiabatic capture.
The particle energy $E$ as a function of time in a ramping cycle is: 
\begin{equation}
\label{eq:E_total}
\begin{split}
 & E(t) = m_0c^2 + K(t), \\
 & K(t) = \left(\frac{K_f-K_i}{2}\right) \left[
           \left(
           \frac{K_f+K_i}{K_f-K_i}
           \right)
           -\cos(2\pi ft)
           \right],
\end{split} 
\end{equation}
where $K(t)$ is the kinetic energy obtained from RF voltage, 
$K_i$ and $K_f$ are initial and final kinetic energies 
for a particle in a ramping cycle, 
$f$ is the booster ramping frequency.
The bending magnetic field for a booster ring 
as a function of time, $B(t)$, should be cooperated with $K(t)$ 
in order to keep particles in the same orbit:
\begin{equation}
\label{eq:B_t}
\begin{split}
    & E(t)^2 = m_0^2c^4 + P(t)^2c^2, \\
    & B(t) = \frac{P(t)}{e\rho} = \frac{1}{e\rho}\sqrt{\frac{E(t)^2-m_0^2c^4}{c^2}}.
\end{split} 
\end{equation}
For the particle with a small stationary mass or very high kinetic energy 
compared to $m_0c^2$, the $B(t)$ is approximately proportional to $K(t)$.
From Eq.~\ref{eq:PhaseSpace} the change of total energy for a particle 
per revolution is equal to the change of kinetic energy:
\begin{equation}
\begin{split}
\label{eq:DE}
  eV\sin\phi_s = E(t_{n+1}) - E(t_{n}) 
               = K(t_{n+1}) - K(t_{n}), 
\end{split}
\end{equation}
where $t_{n+1}-t_n$ is the revolution period for a particle between 
$n$ and $n+1$ turns:
\begin{equation}
\begin{split}
\label{eq:Dt}
  t_{n+1}-t_n = \frac{L}{v(E(t_n))}.
\end{split}
\end{equation}
$L$ is the circumference for a ring accelerator, 
and the particle velocity $v$ is a function 
of particle energy at the time $t_n$, as shown in Eq.~\ref{eq:PhaseSpace}.
The RF phase factor $\phi_s$ is thus calculated by
\begin{equation}
\label{eq:phi_s}
  \phi_s = \sin^{-1}\left[\frac{K(t_{n+1})-K(t_{n})}{eV}\right].
\end{equation}
Note that this equation also sets the maximum rate for 
the ramping of kinetic energy with $0 \le [K(t_{n+1})-K(t_{n})] < eV$. 
For a heavy particle ramping, if it is started at low kinetic energy, 
the particle velocity is not close to the speed of light.
So the revolution period would be changed obviously 
at the early stage of ramping.
Since $\phi_s$ is dependent on the change of kinetic energy 
per revolution period, it is also changed obviously, 
as shown in the top left and right of Fig.~\ref{fig-proton-ramp}.

The synchrotron tune $Q_s$ is calculated by 
\begin{equation}
\begin{split}
\label{eq:Q_s}
  Q_s = \nu_s\sqrt{|\cos\phi_s|},
  \mathrm{\ \ where\ \ } \nu_s = \sqrt{\frac{h|\eta|eV}{2\pi\beta^2 E}},
\end{split}
\end{equation}
if $\eta \ne 0$.
The adiabatic coefficient $\alpha_{\rm{ad}}$ is then defined as 
\begin{equation}
\begin{split}
\label{eq:adiabatic}
  \alpha_{\rm{ad}} = \frac{1}{2\pi}\left| \frac{dT_s}{dt} \right|,
  \mathrm{\ \ where\ \ } T_s = T_0 / Q_s,
\end{split}
\end{equation}
$T_0$ is the revolution period. 
The condition for adiabatic synchrotron motion is $\alpha_{\rm{ad}} \ll 1$.
The phase-space area enclosed by the separatrix is called bucket area, 
$\tilde{\mathcal{A}}_{\rm{B}}$, which is approximated as 
\begin{equation}
\label{eq:area}
  \tilde{\mathcal{A}}_{\rm{B}} \approx \frac{16Q_s}{h|\eta|\sqrt{|\cos\phi_s|}}
  \left(\frac{1-\sin\phi_s}{1+\sin\phi_s}\right).
\end{equation}
Since bucket area is the maximum size of a beam we can make, 
we should avoid the zero bucket area with $\phi_s=90^{\circ}$ 
from Eq.~\ref{eq:phi_s}.

\section{OPTIMIZE THE TLS BOOSTER AS A PROTON OR ELECTRON MACHINE}
\begin{figure}[htb]
    \centering
    \epsfig{file=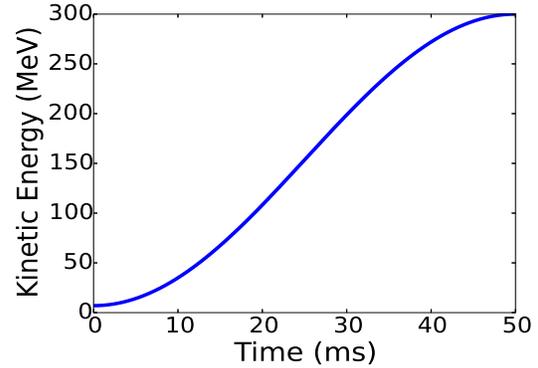,width=70mm,height=50mm}
    \caption{The particle kinetic energy as a function of time 
             in a ramping cycle, where $K_i=7$~(MeV), 
             $K_f=300$~(MeV) and $f=10$~(Hz).}
    \label{fig-kinetic}
\end{figure}

For the TLS booster, the fixed parameters are 
$L=72$ (m), $\rho=5$ (m), $\alpha_c=0.1346$,
$K_i=7$~(MeV), $K_f=300$~(MeV) and $f=10$~(Hz).
For RF cavity, the fixed parameters are $h=2$ and $V_f=15$~(kV). 
The kinetic energy as a function of time for particles 
in a ramping cycle is shown in Fig.~\ref{fig-kinetic}, 
which is based on Eq.~\ref{eq:E_total}.
The RF voltage variables $T_{\nu}$ and $V_i$ are crucial parameters to be 
optimized for the best efficiency of adiabatic capture.
We assume the initial conditions for a bunch of the beam is 
flatly distributed in the phase $\phi=[-\pi,~\pi]$ or $\phi=[0,~2\pi]$ (rad), 
and the distribution of $\Delta P/P$ ($\Delta E/E$) for 
protons (electrons) is a Gaussian with zero mean 
and $\sigma=\pm0.05\%$ or $\sigma=\pm0.5\%$, i.e., 
the width of $\Delta P/P$ ($\Delta E/E$) is 0.1\% or 1\% 
for protons (electrons).
Two thousand particles are generated to represent a bunch of the beam 
($N=2000$). 
We track these particles by the synchrotron motion simulator and obtain 
the capture efficiency for particles in a ramping cycle.

\begin{figure}[htb]
    \centering
    \epsfig{file=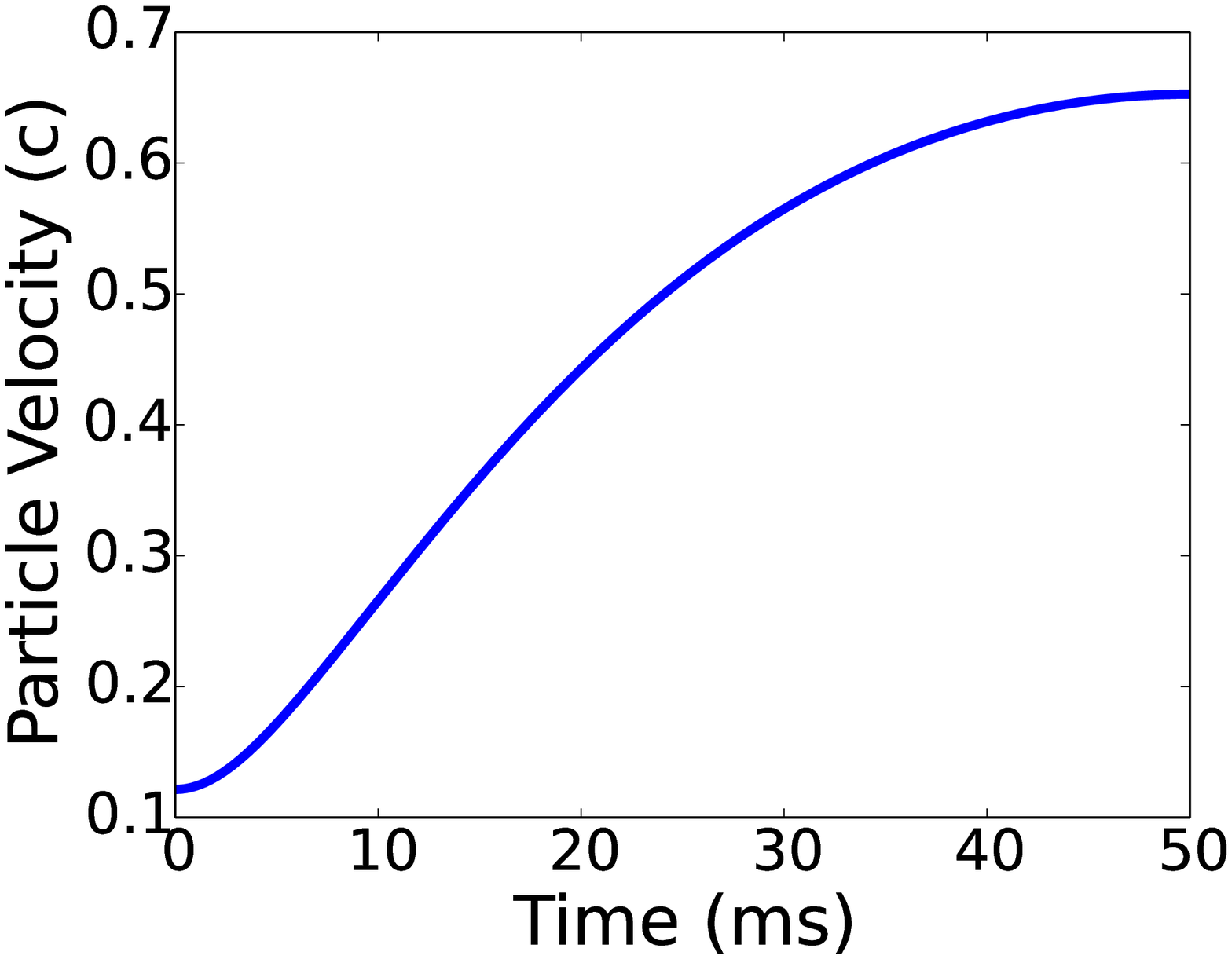,width=40mm,height=38mm}
    \epsfig{file=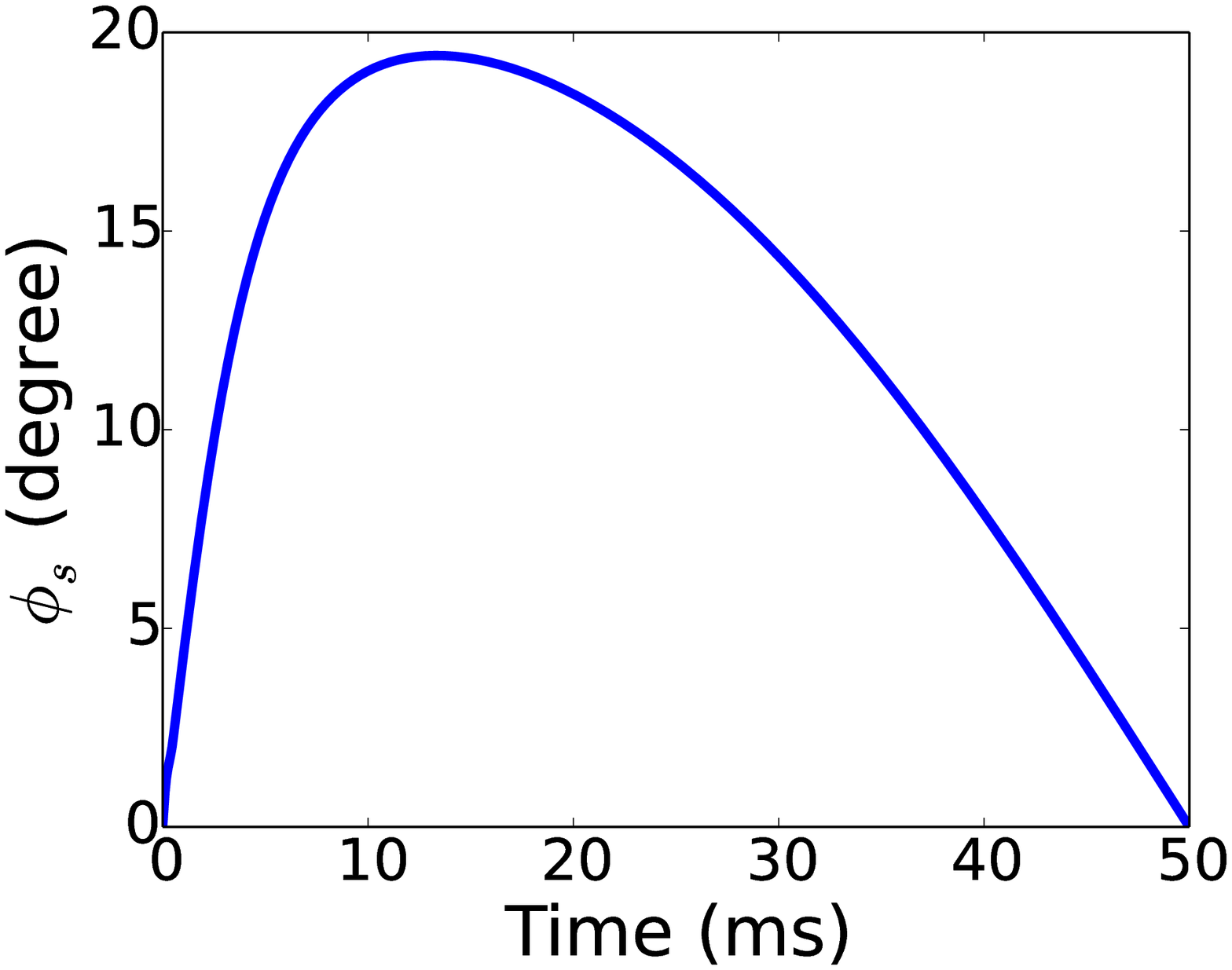,width=40mm,height=38mm}
    \epsfig{file=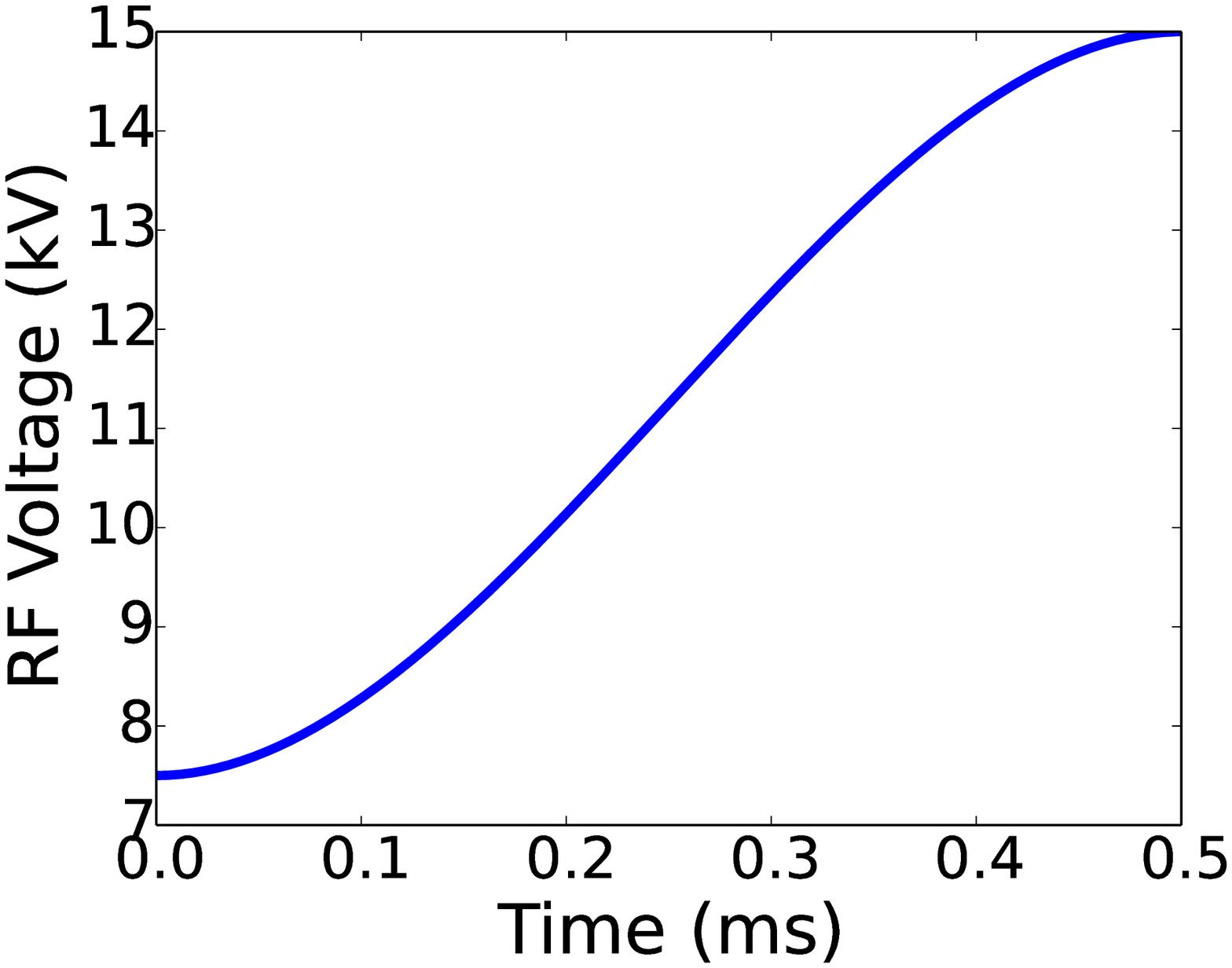,width=40mm,height=38mm}
    \epsfig{file=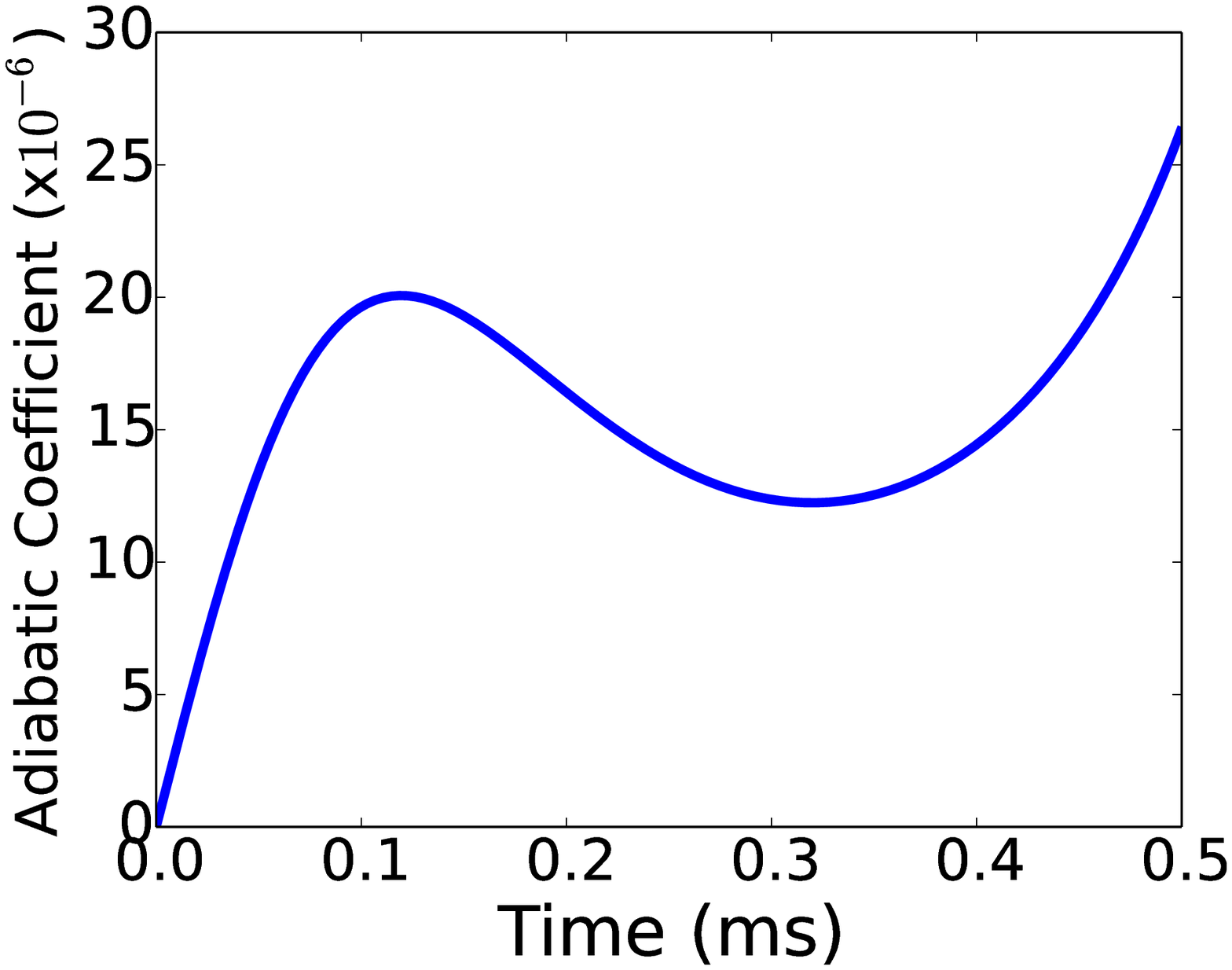,width=40mm,height=38mm}
    \epsfig{file=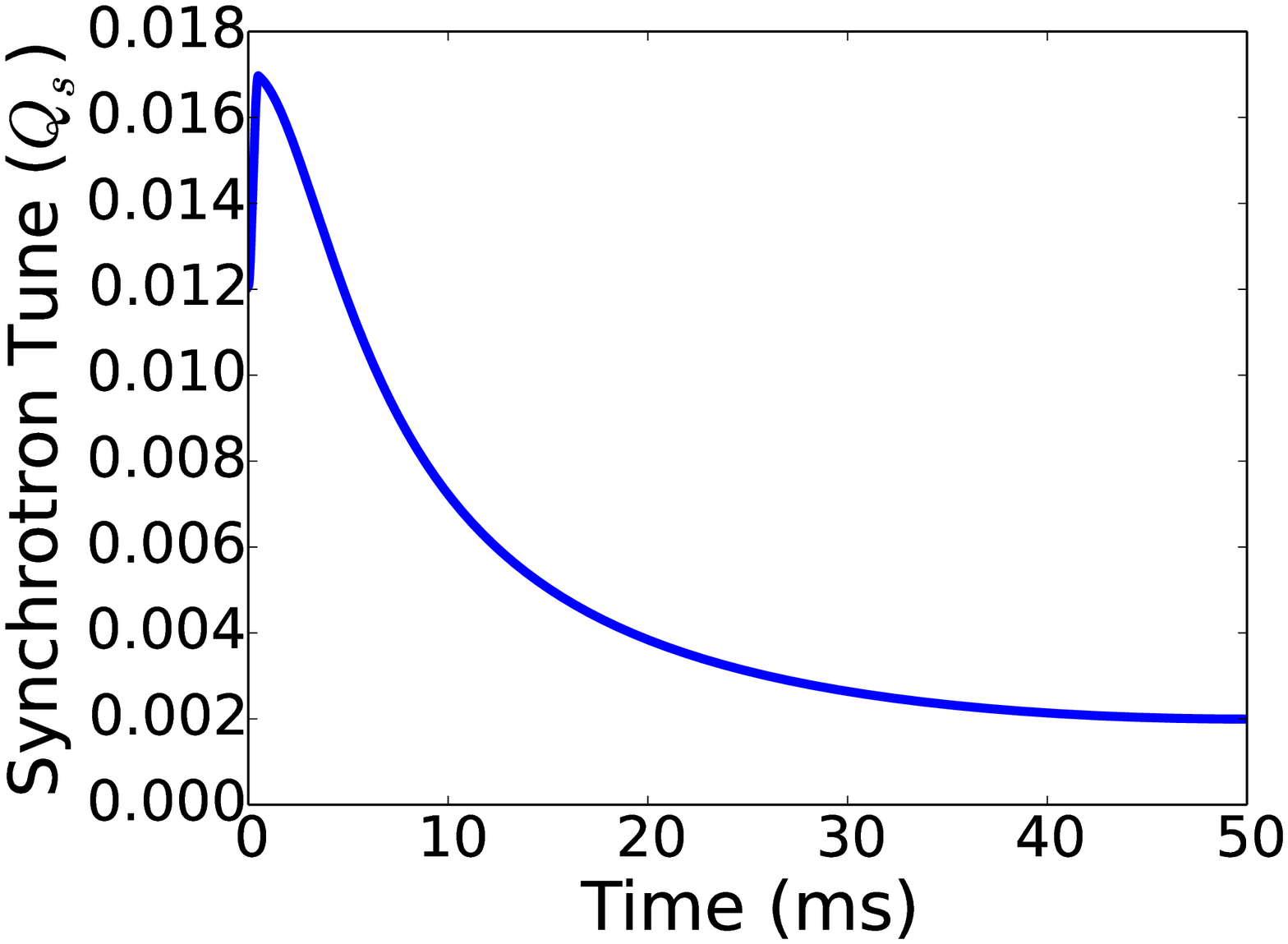,width=40mm,height=38mm}
    \epsfig{file=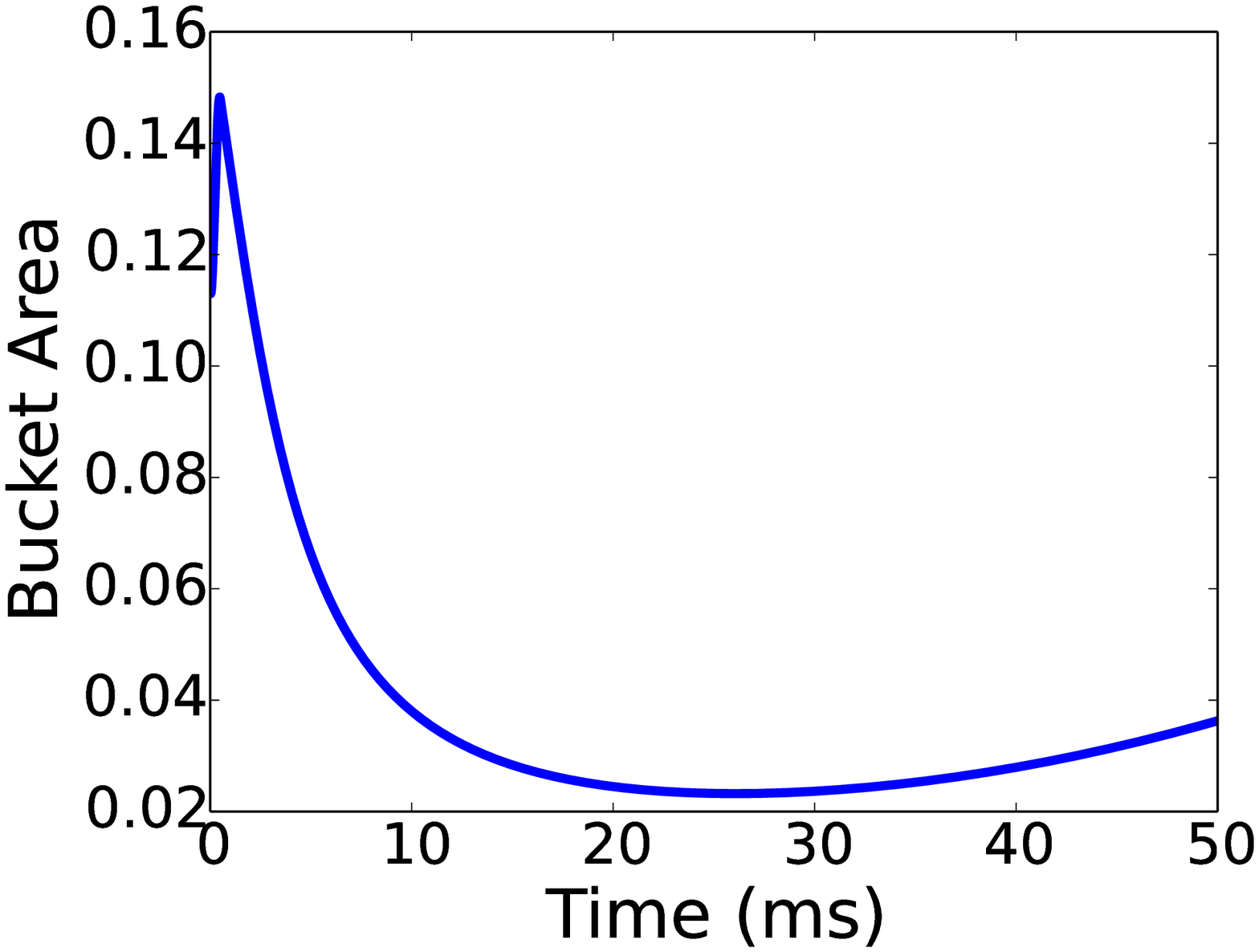,width=40mm,height=38mm}
    \caption{The proton velocity in the unit of light speed $c$ (top left), 
             RF phase factor $\phi_s$ (top right), 
             RF voltage $V$ (middle left), 
             adiabatic coefficient $\alpha_{\rm{ad}}$ (middle right), 
             synchrotron tune $Q_s$ (bottom left) and 
             bucket area $\tilde{\mathcal{A}}_{\rm{B}}$ (bottom right) 
             vs. time in a ramping cycle.
             Here we set $T_{\nu}=0.5$ (ms) and $V_i=7.5$ (kV).}
    \label{fig-proton-ramp}
\end{figure}

\begin{figure}[htb]
    \centering
    \epsfig{file=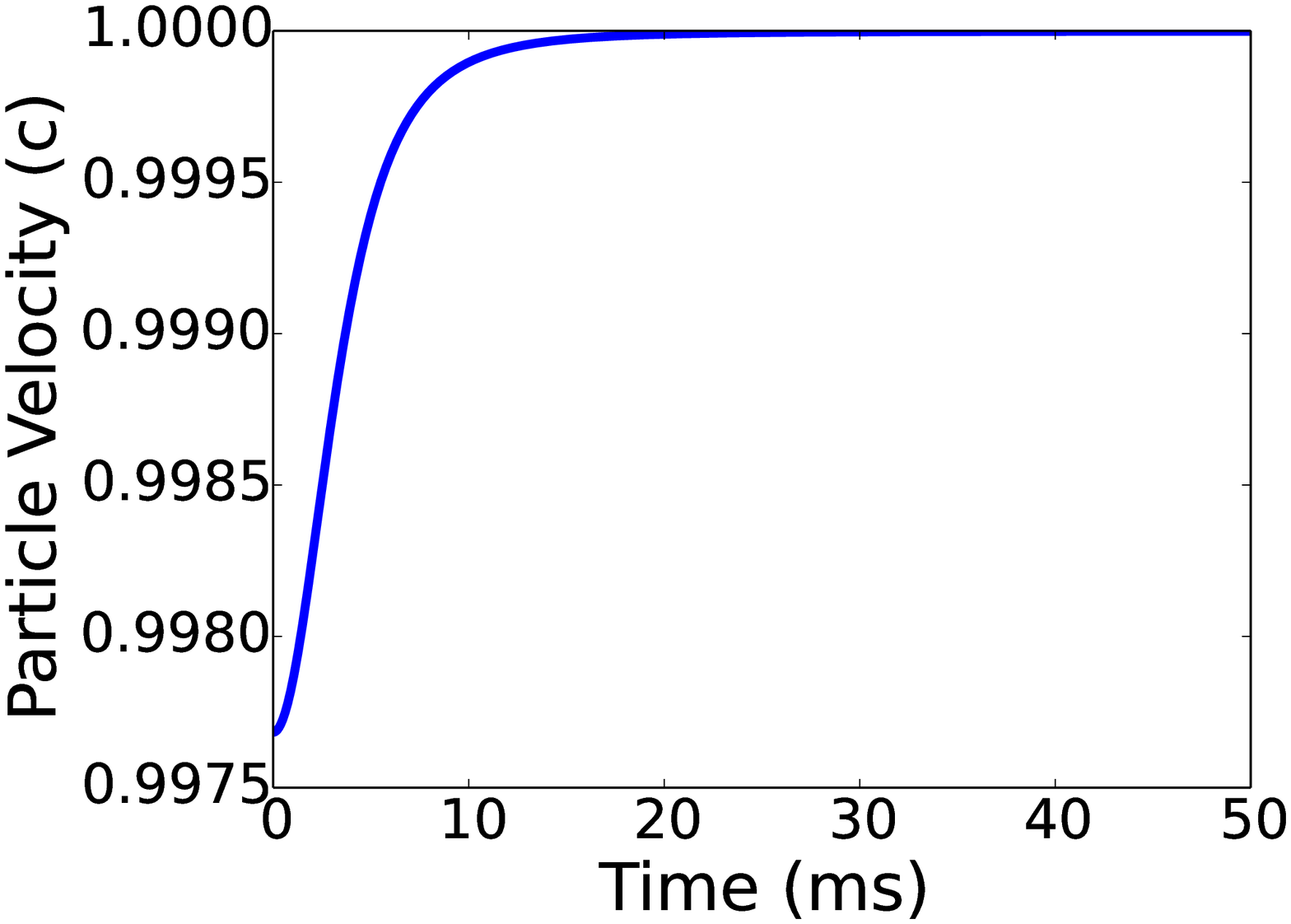,width=40mm,height=38mm}
    \epsfig{file=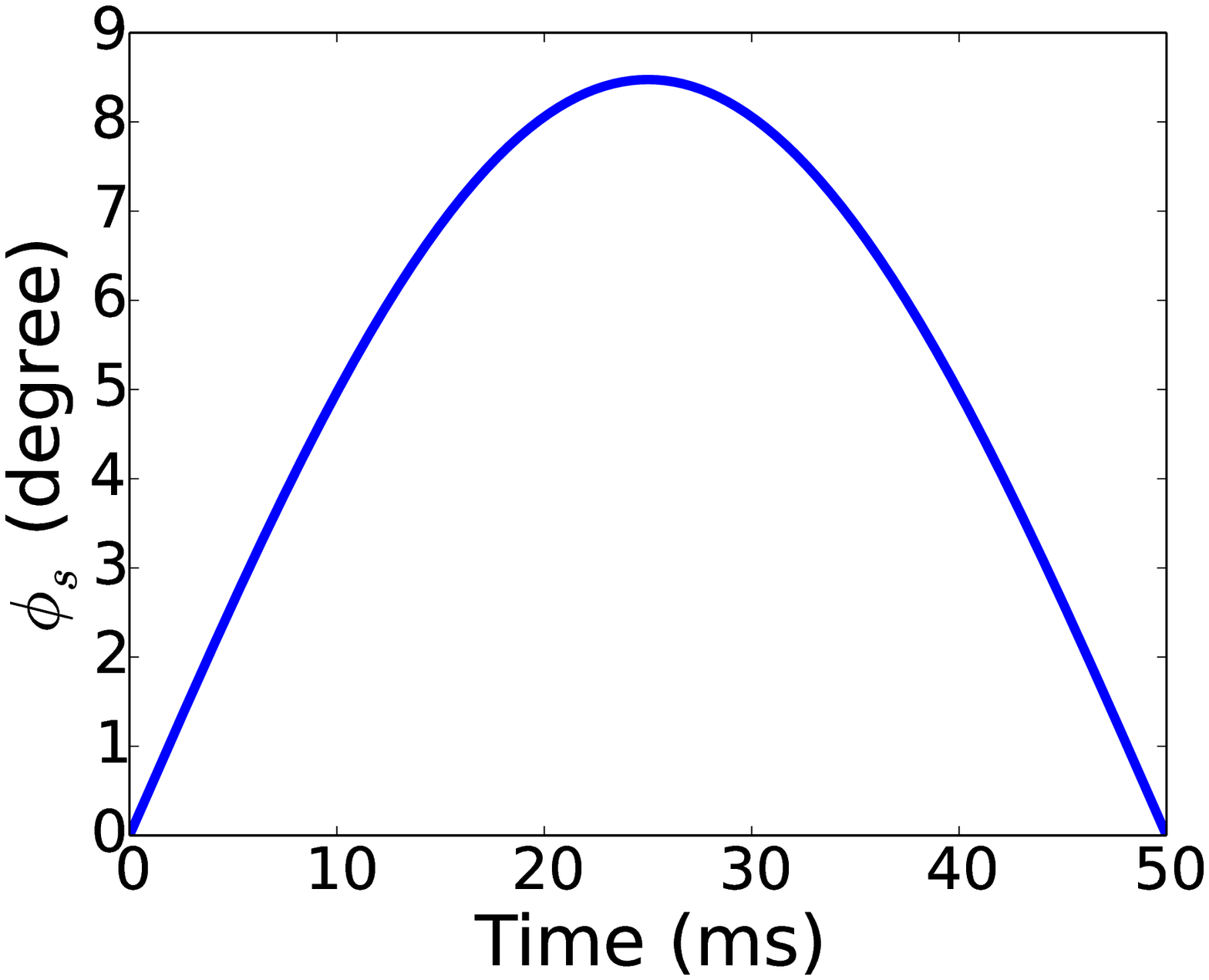,width=40mm,height=38mm}
    \epsfig{file=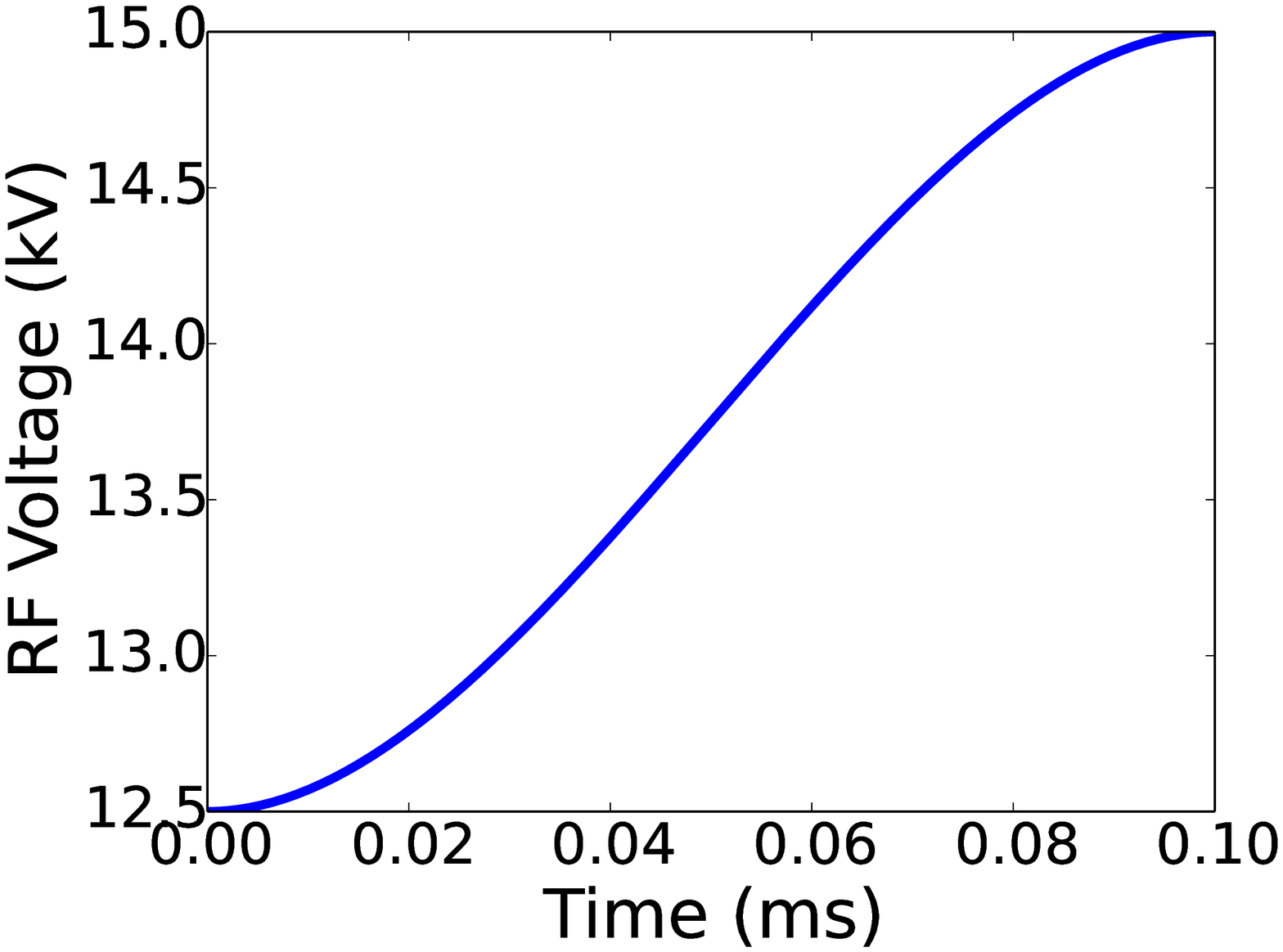,width=40mm,height=38mm}
    \epsfig{file=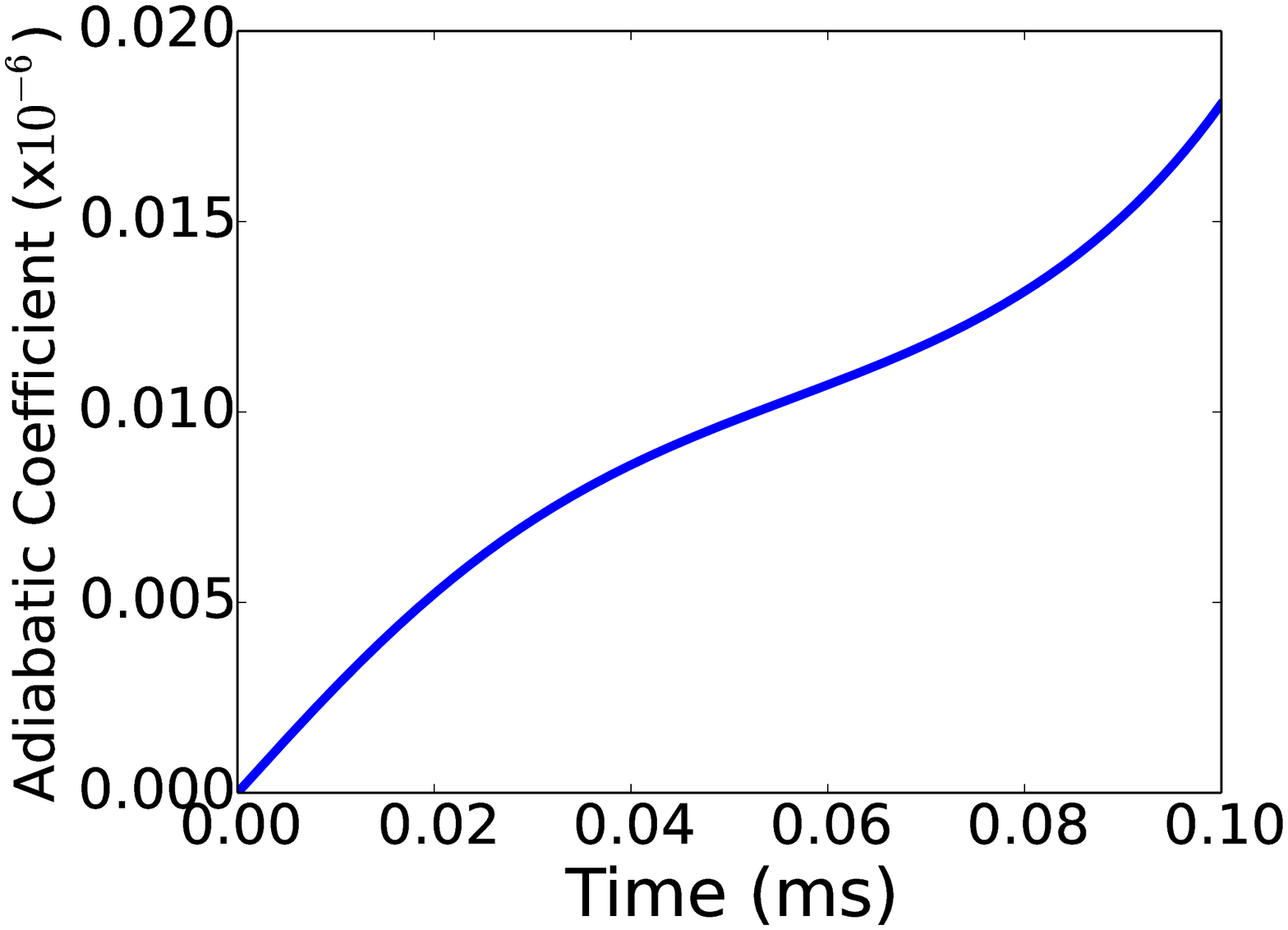,width=40mm,height=38mm}
    \epsfig{file=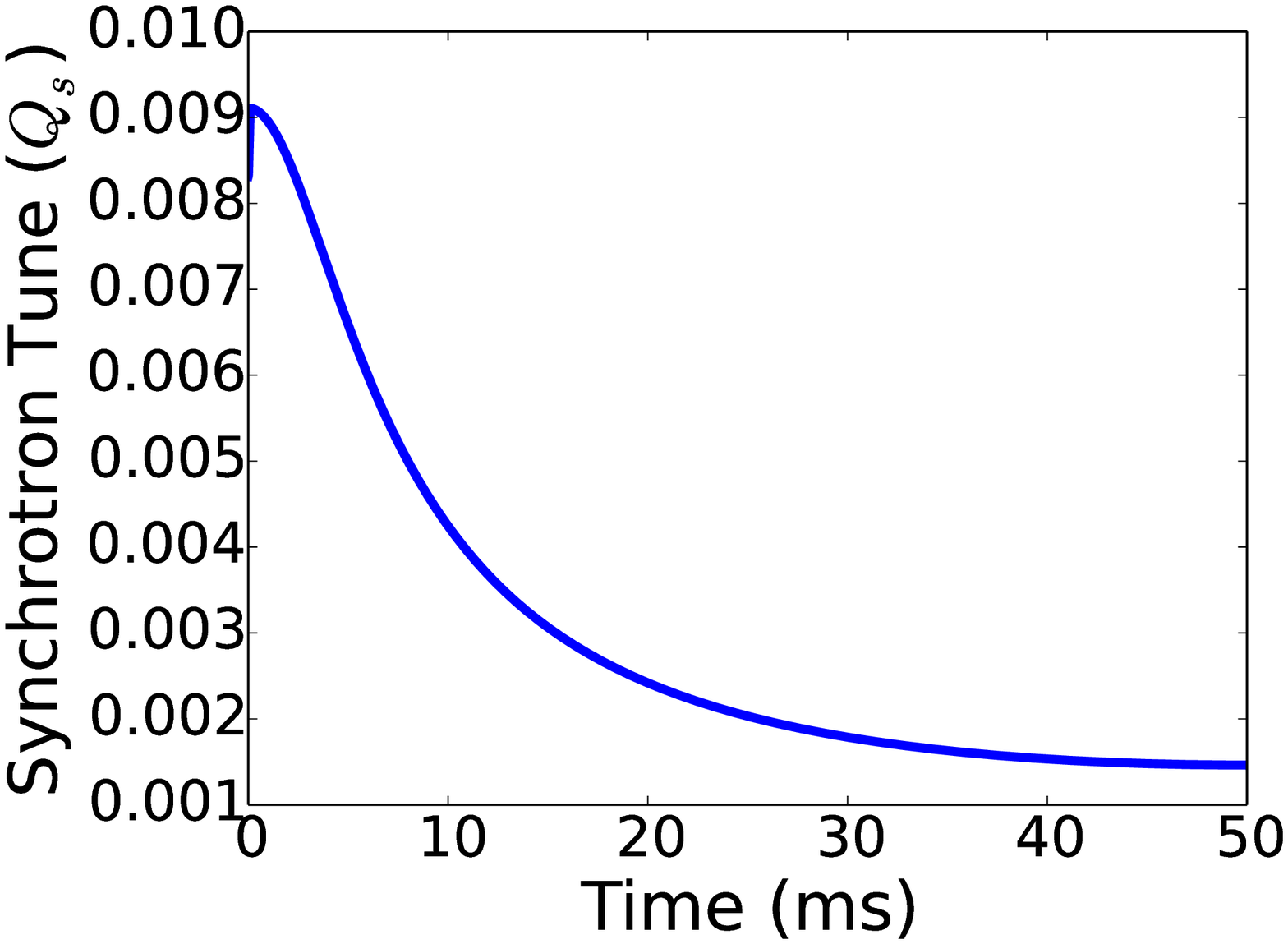,width=40mm,height=38mm}
    \epsfig{file=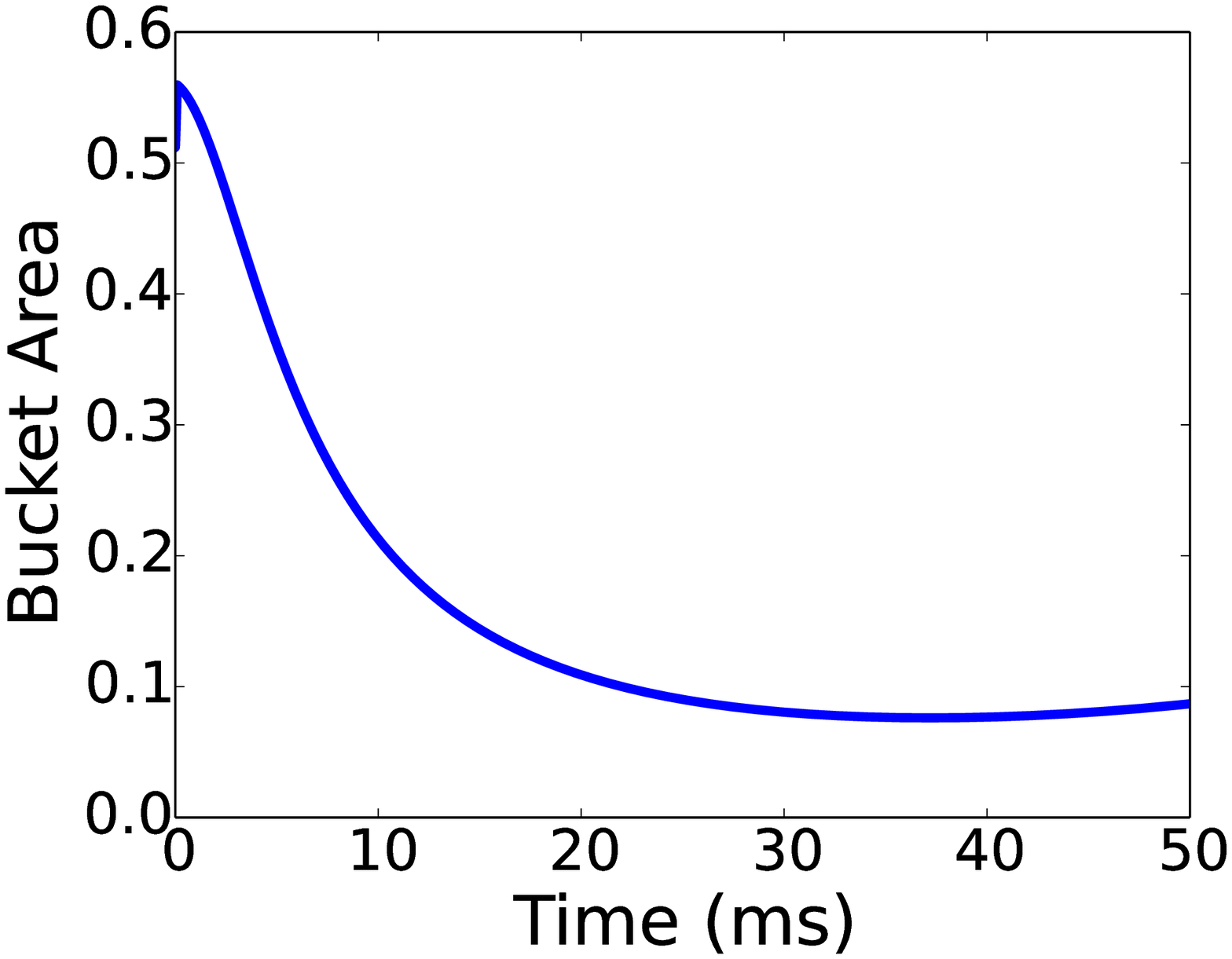,width=40mm,height=38mm}
    \caption{The electron velocity in the unit of light speed $c$ (top left), 
             RF phase factor $\phi_s$ (top right), 
             RF voltage $V$ (middle left), 
             adiabatic coefficient $\alpha_{\rm{ad}}$ (middle right), 
             synchrotron tune $Q_s$ (bottom left) and  
             bucket area $\tilde{\mathcal{A}}_{\rm{B}}$ (bottom right) 
             vs. time in a ramping cycle.
             Here we set $T_{\nu}=0.1$ (ms) and $V_i=12.5$ (kV).}
    \label{fig-electron-ramp}
\end{figure}

\begin{figure}[htb]
    \centering
    \epsfig{file=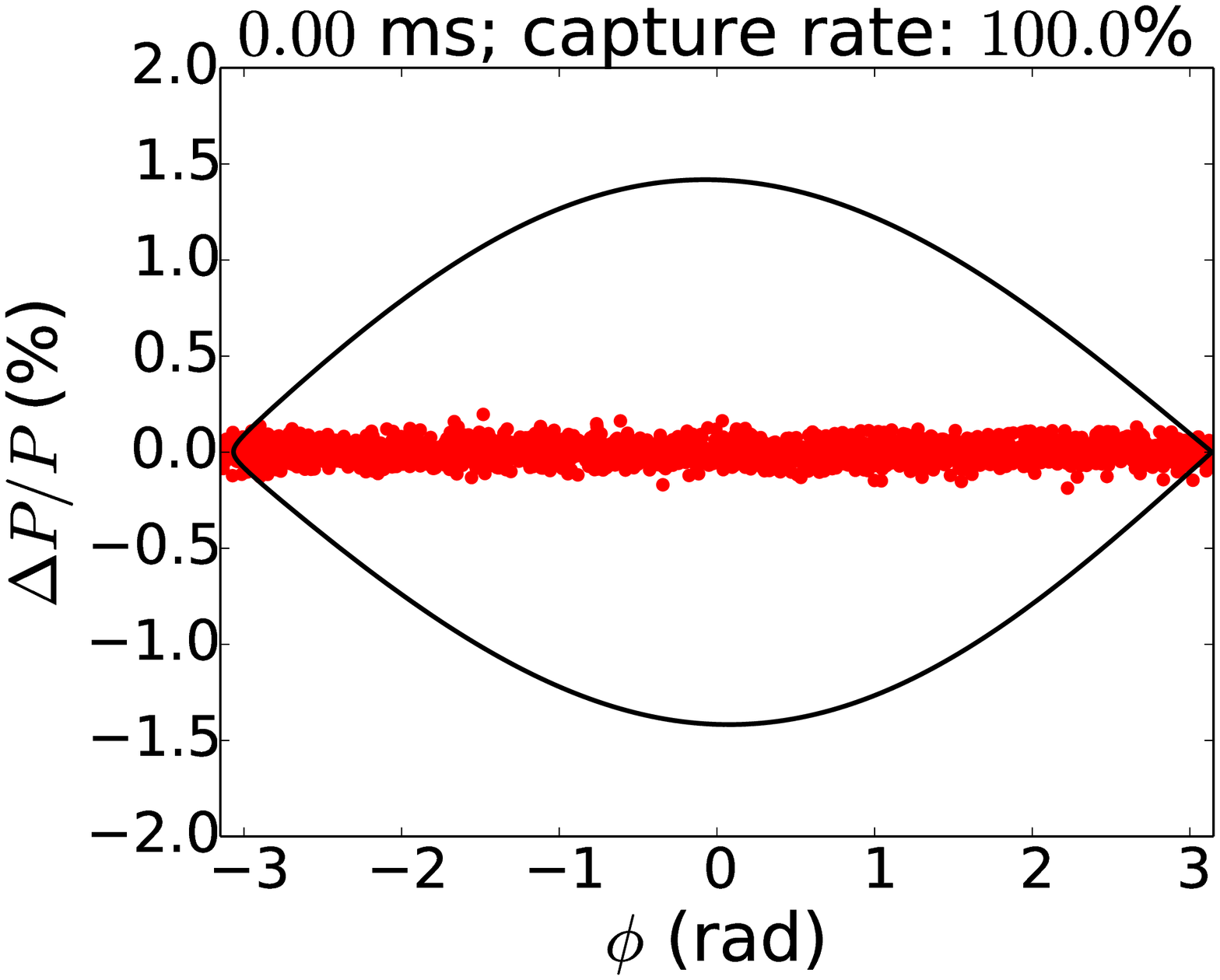,width=40mm,height=40mm}
    \epsfig{file=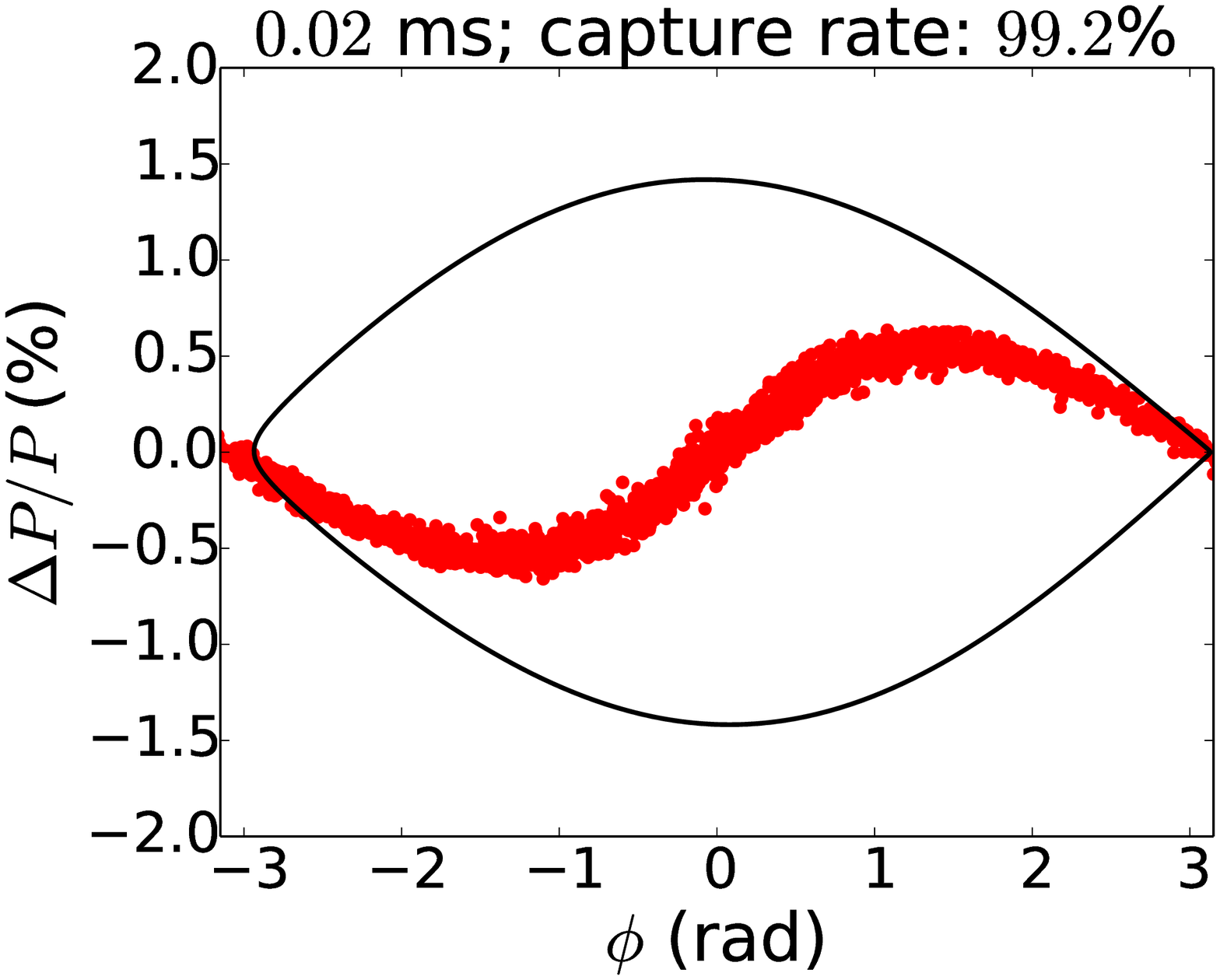,width=40mm,height=40mm}
    \epsfig{file=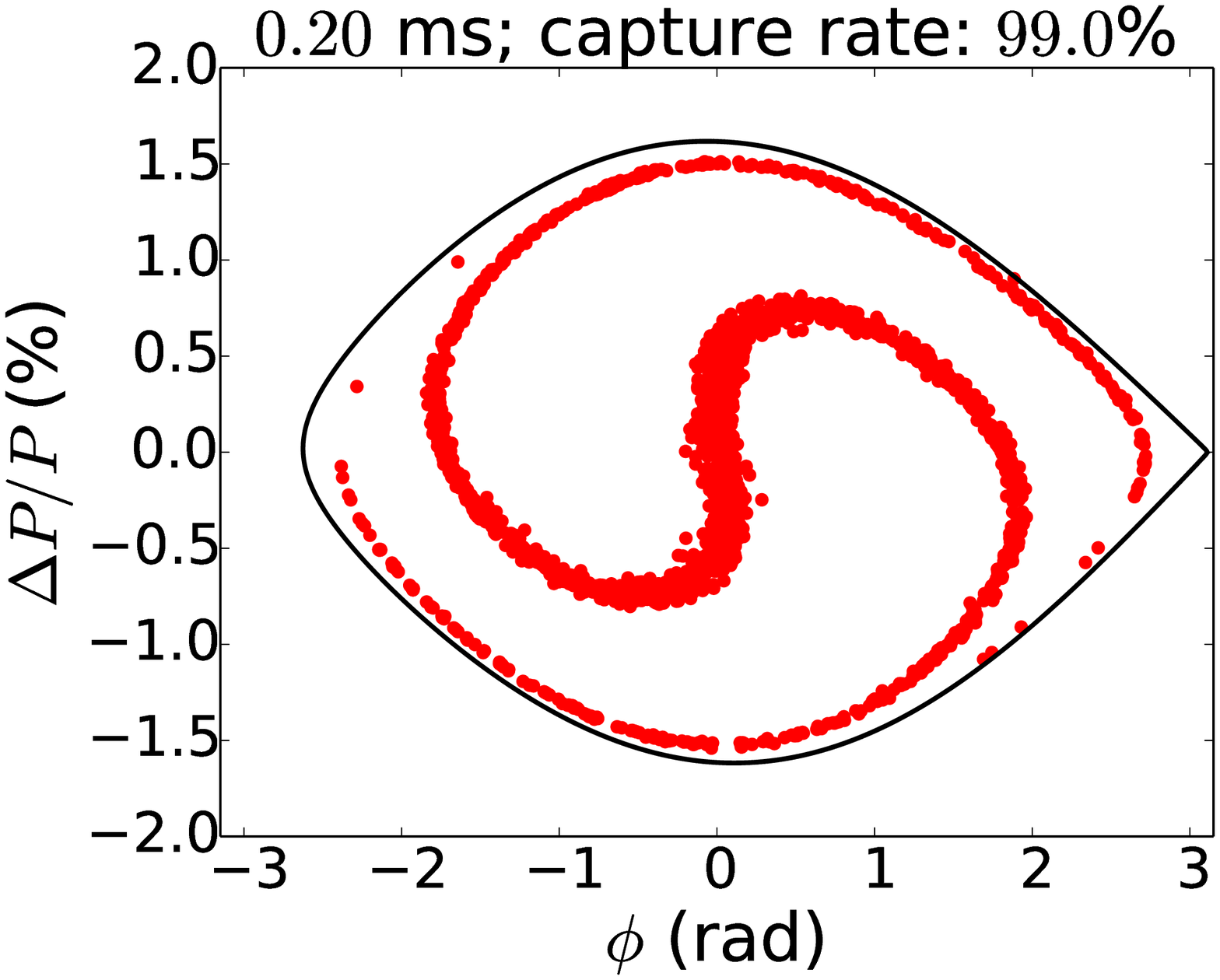,width=40mm,height=40mm}
    \epsfig{file=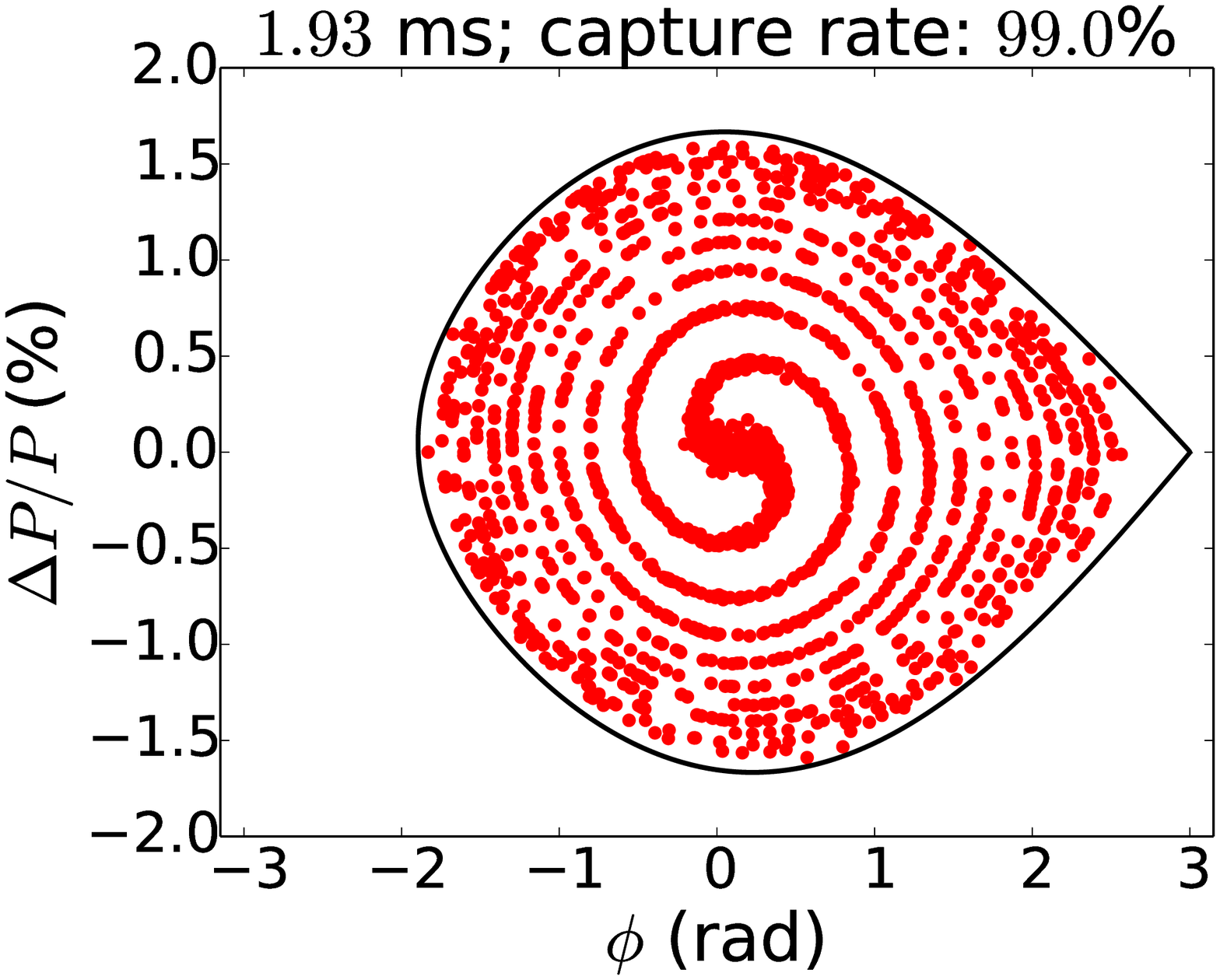,width=40mm,height=40mm}
    \epsfig{file=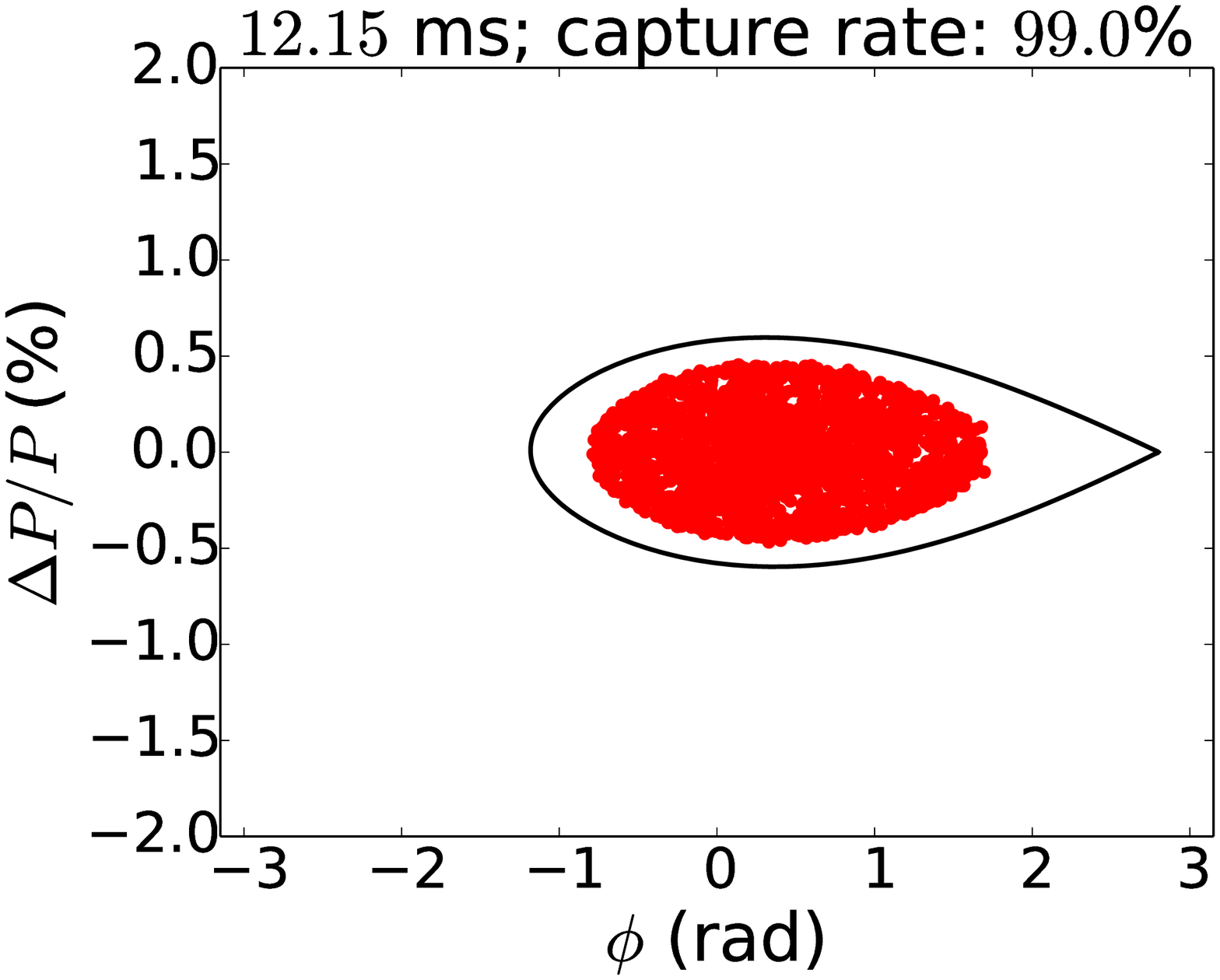,width=40mm,height=40mm}
    \epsfig{file=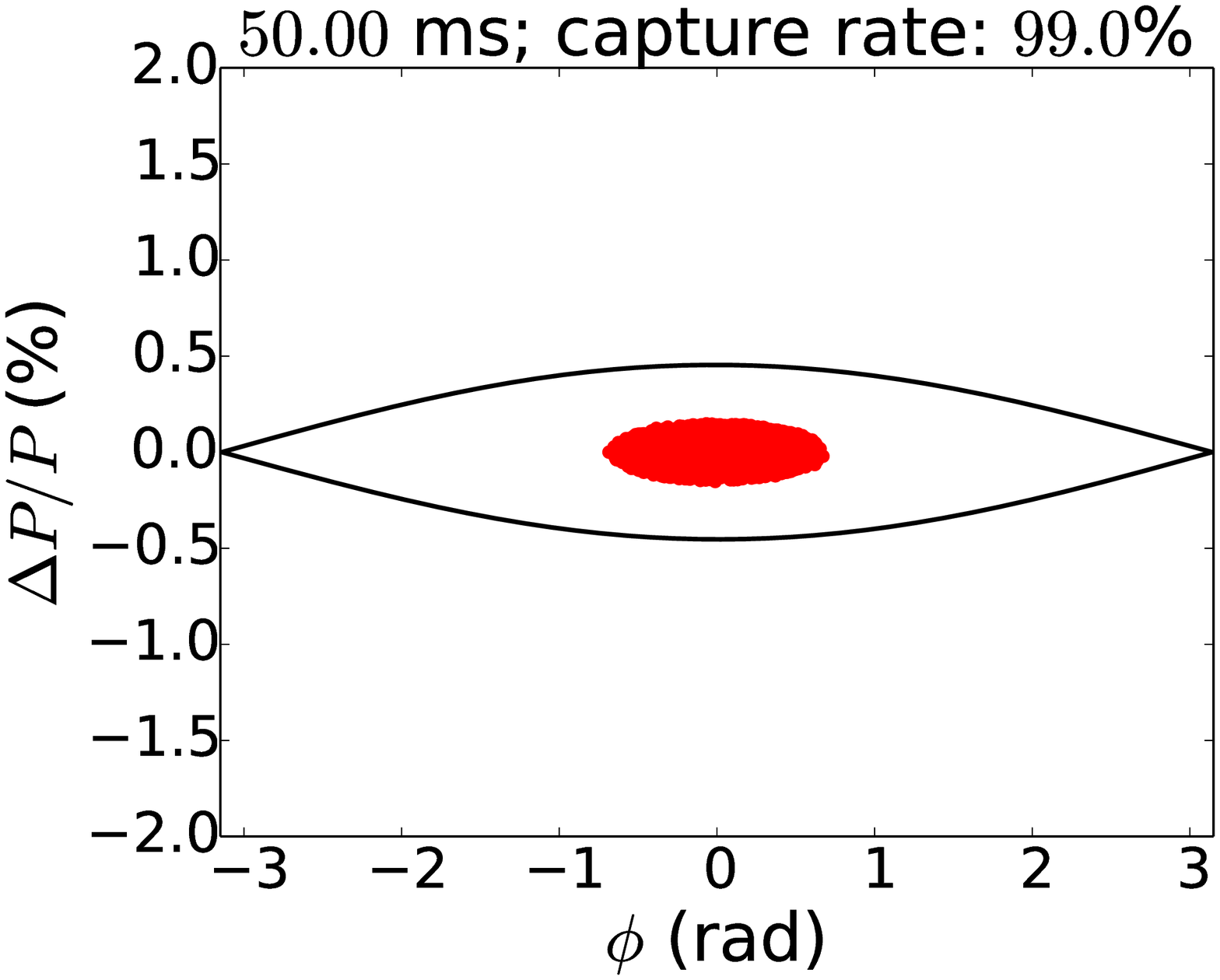,width=40mm,height=40mm}    
    \epsfig{file=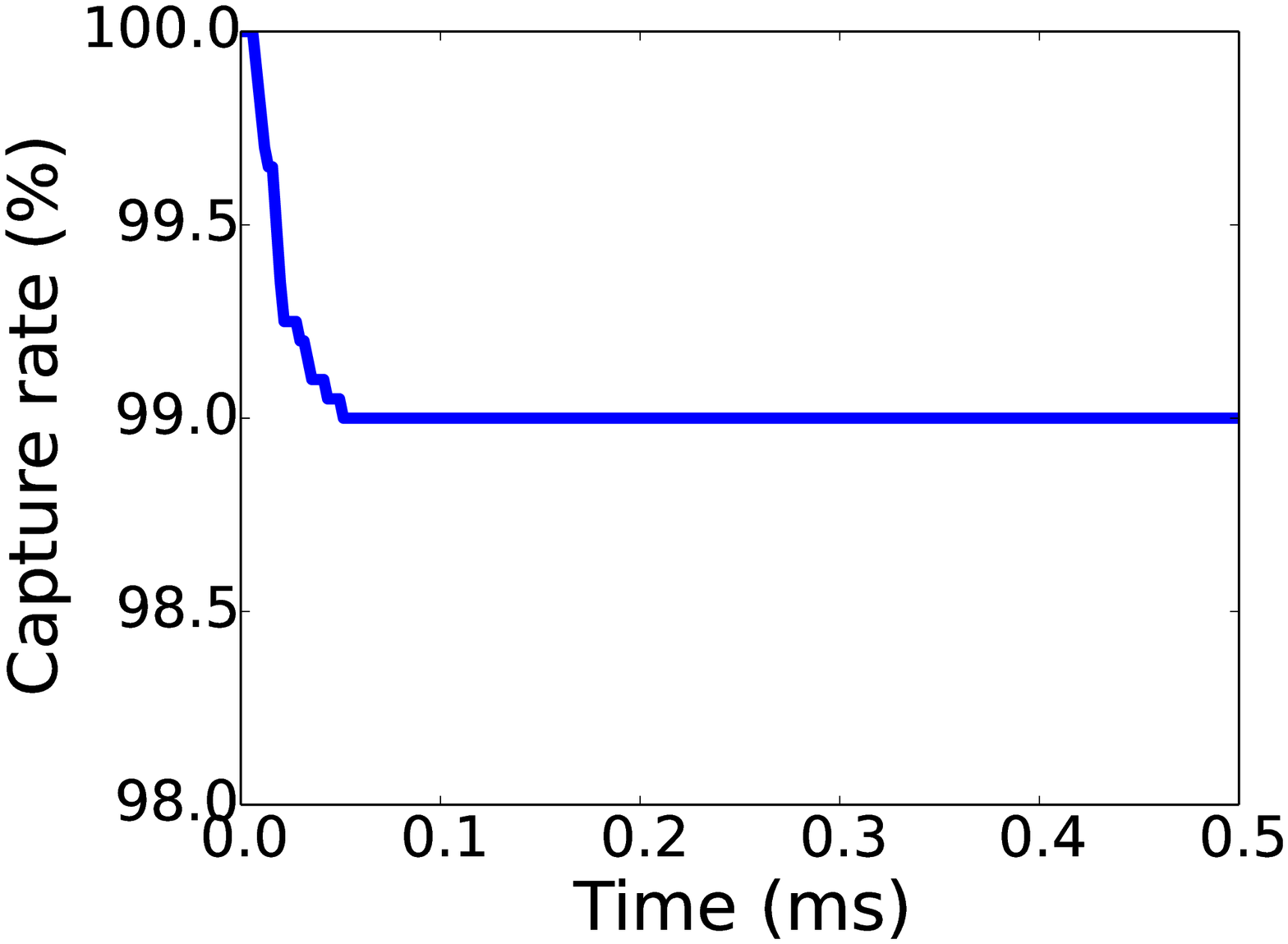,width=60mm,height=45mm}
    \caption{The evolution of phase-space $(\phi,\Delta P/P)$ for 
             protons in a ramping cycle. 
             We set $T_{\nu}=0.5$ (ms), $V_i=7.5$ (kV) and 
             assume the initial conditions for a bunch of the beam is 
             flatly distributed in the phase $\phi=[-\pi,~\pi]$ (rad), 
             and the $\Delta P/P$ is a Gaussian distribution with zero mean 
             and $\sigma=\pm0.05\%$, i.e., the width of $\Delta P/P$ is 0.1\%.
             Bottom plot is the capture rate vs. time for protons in a ramping cycle.}
    \label{fig-proton-phase}
\end{figure}

\begin{figure}[htb]
    \centering
    \epsfig{file=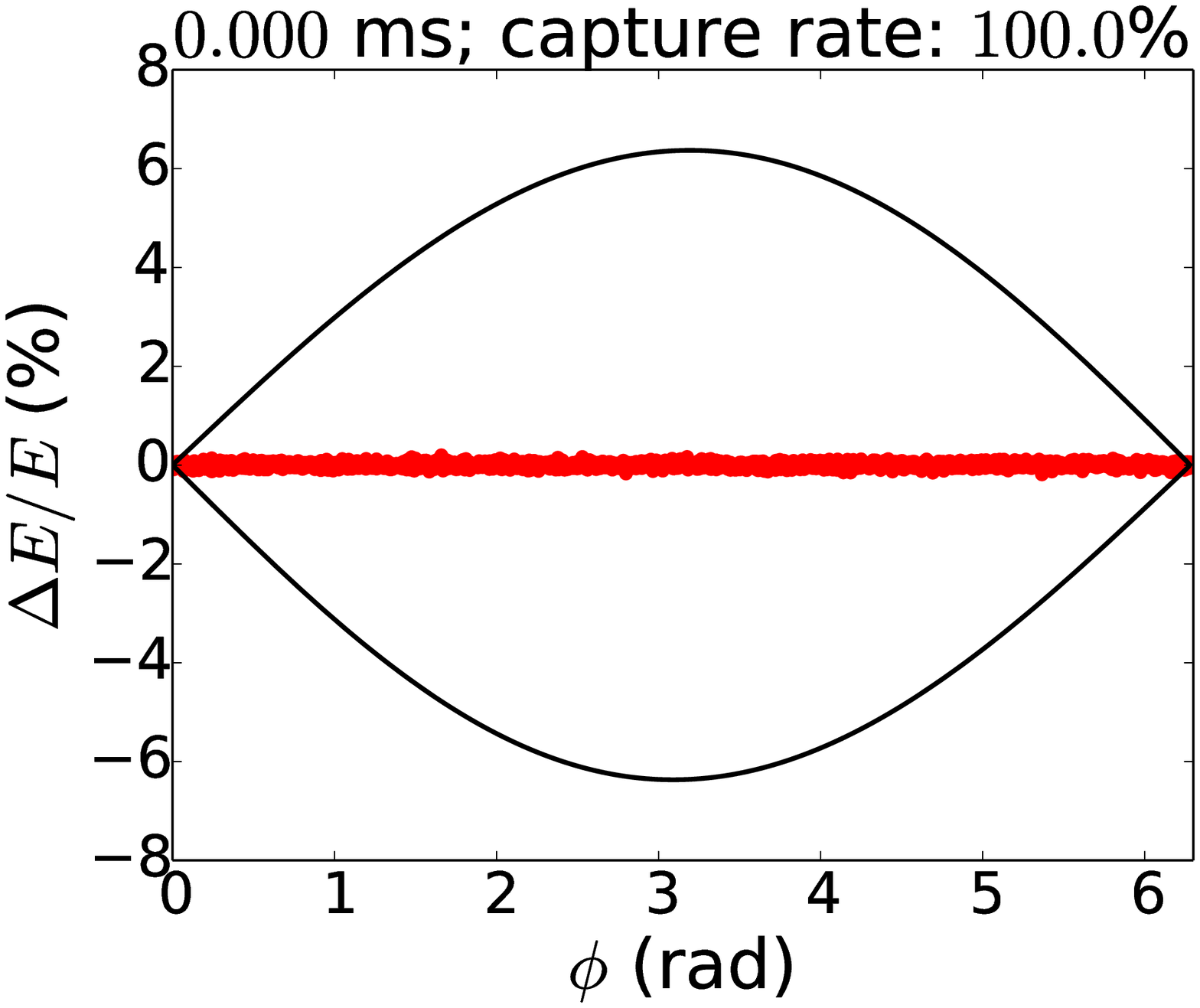,width=40mm,height=40mm}
    \epsfig{file=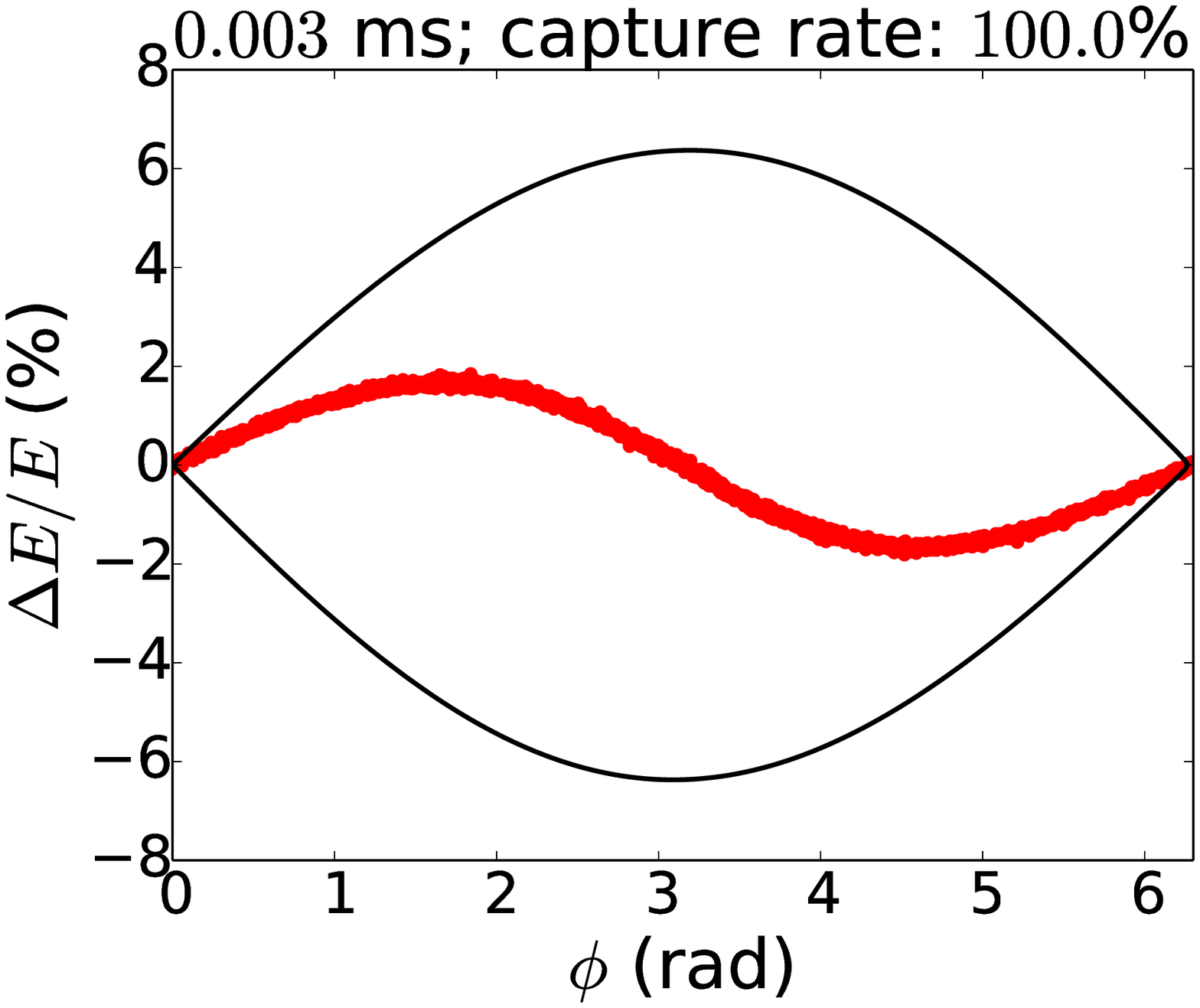,width=40mm,height=40mm}
    \epsfig{file=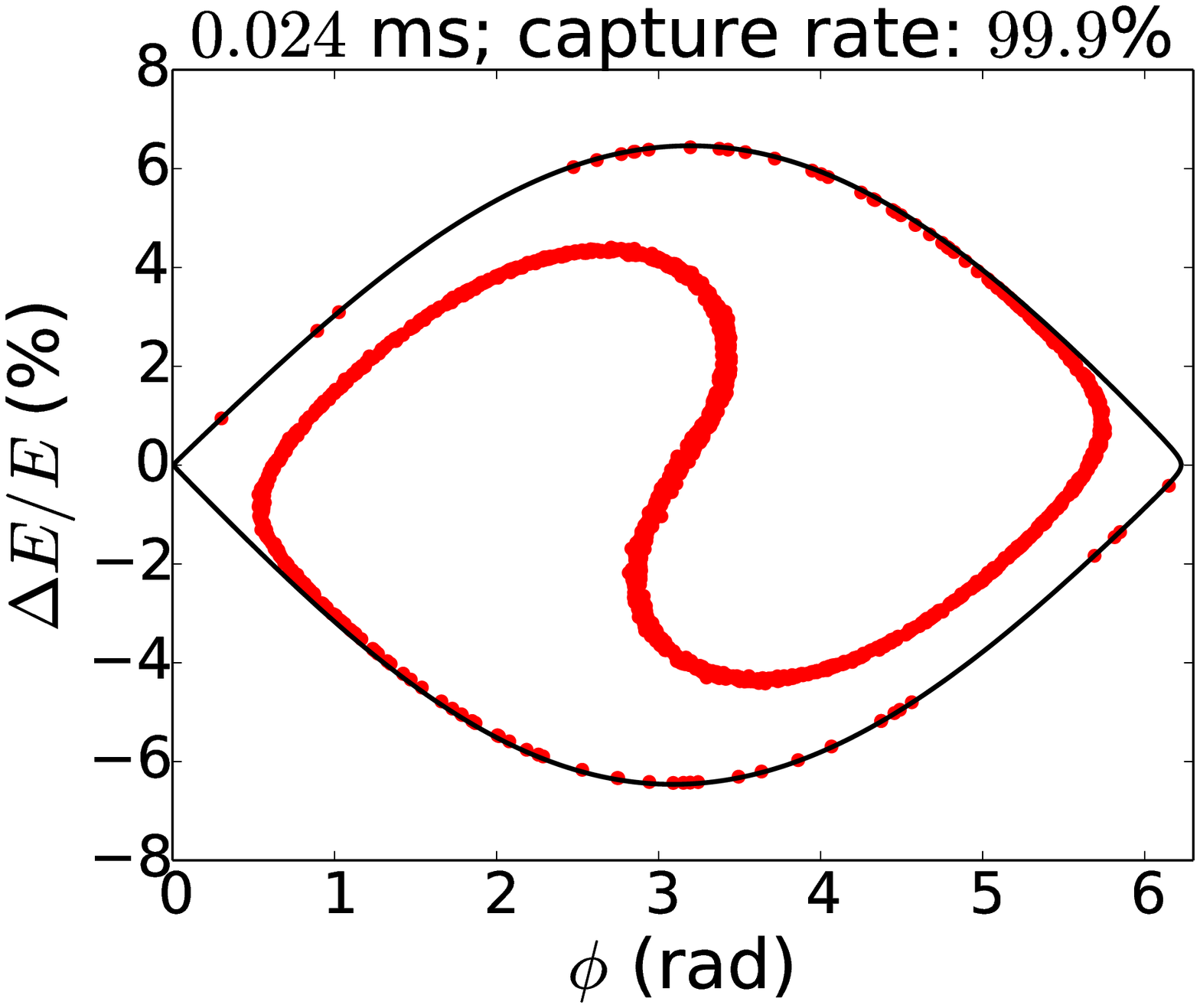,width=40mm,height=40mm}
    \epsfig{file=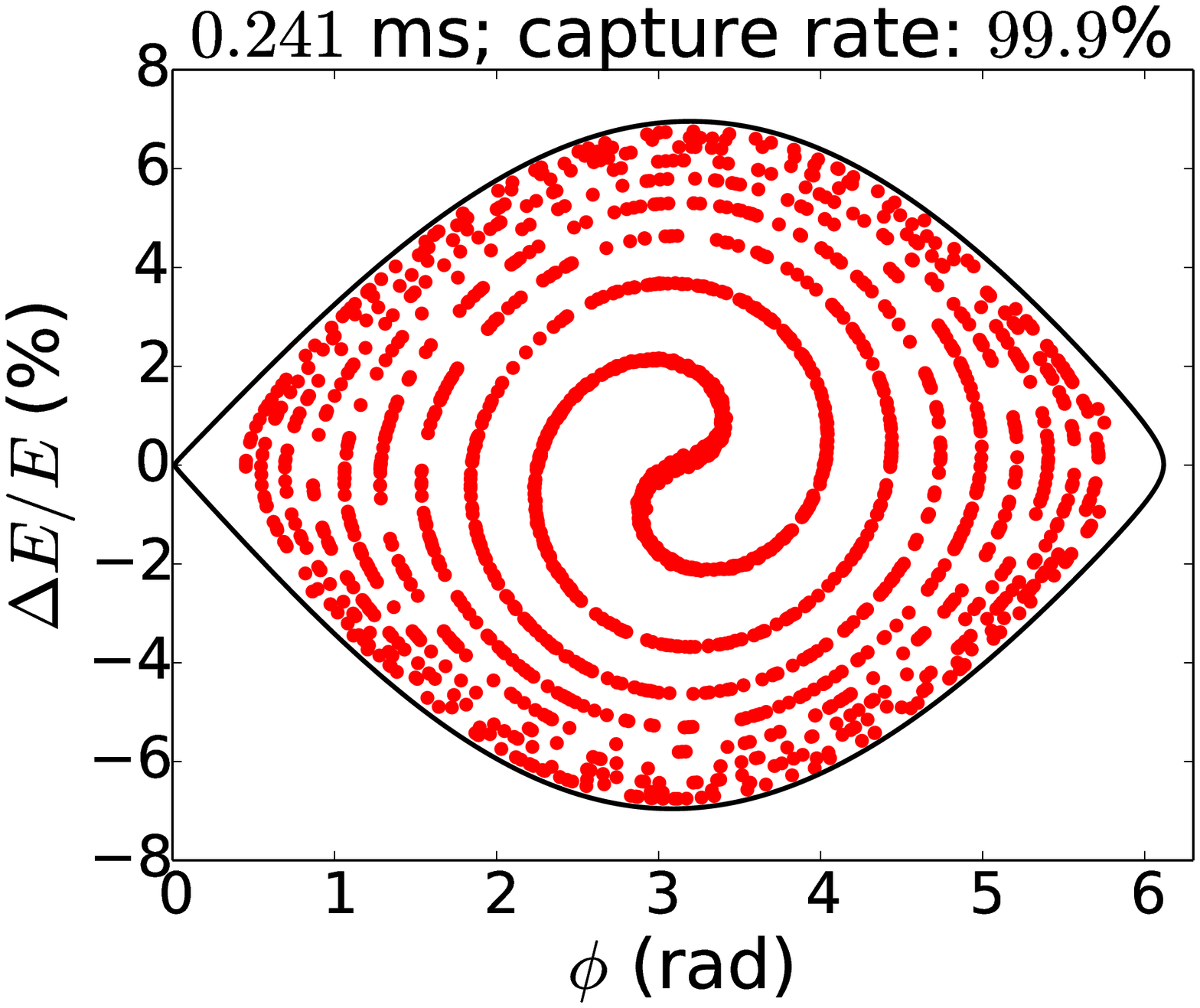,width=40mm,height=40mm}
    \epsfig{file=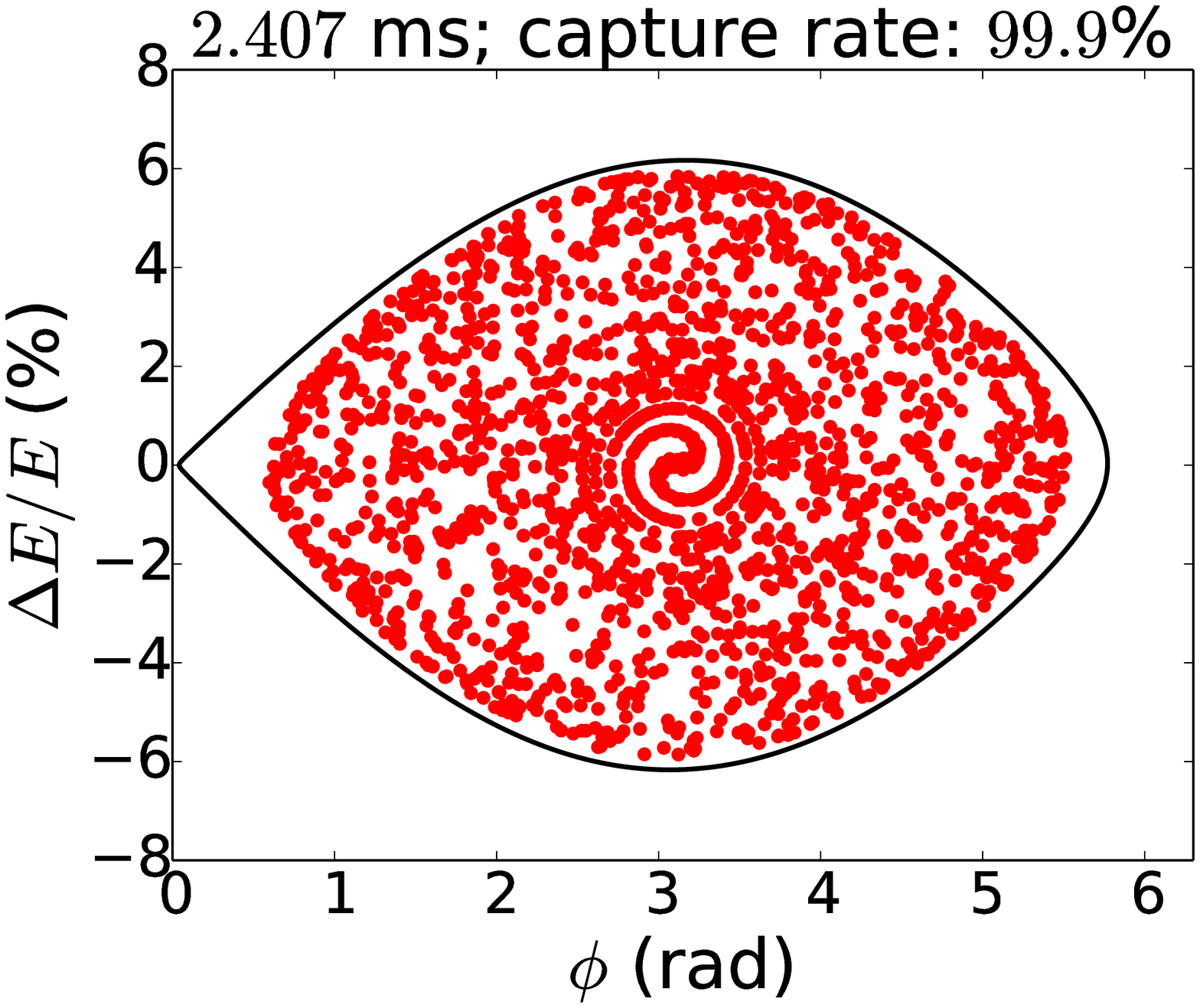,width=40mm,height=40mm}
    \epsfig{file=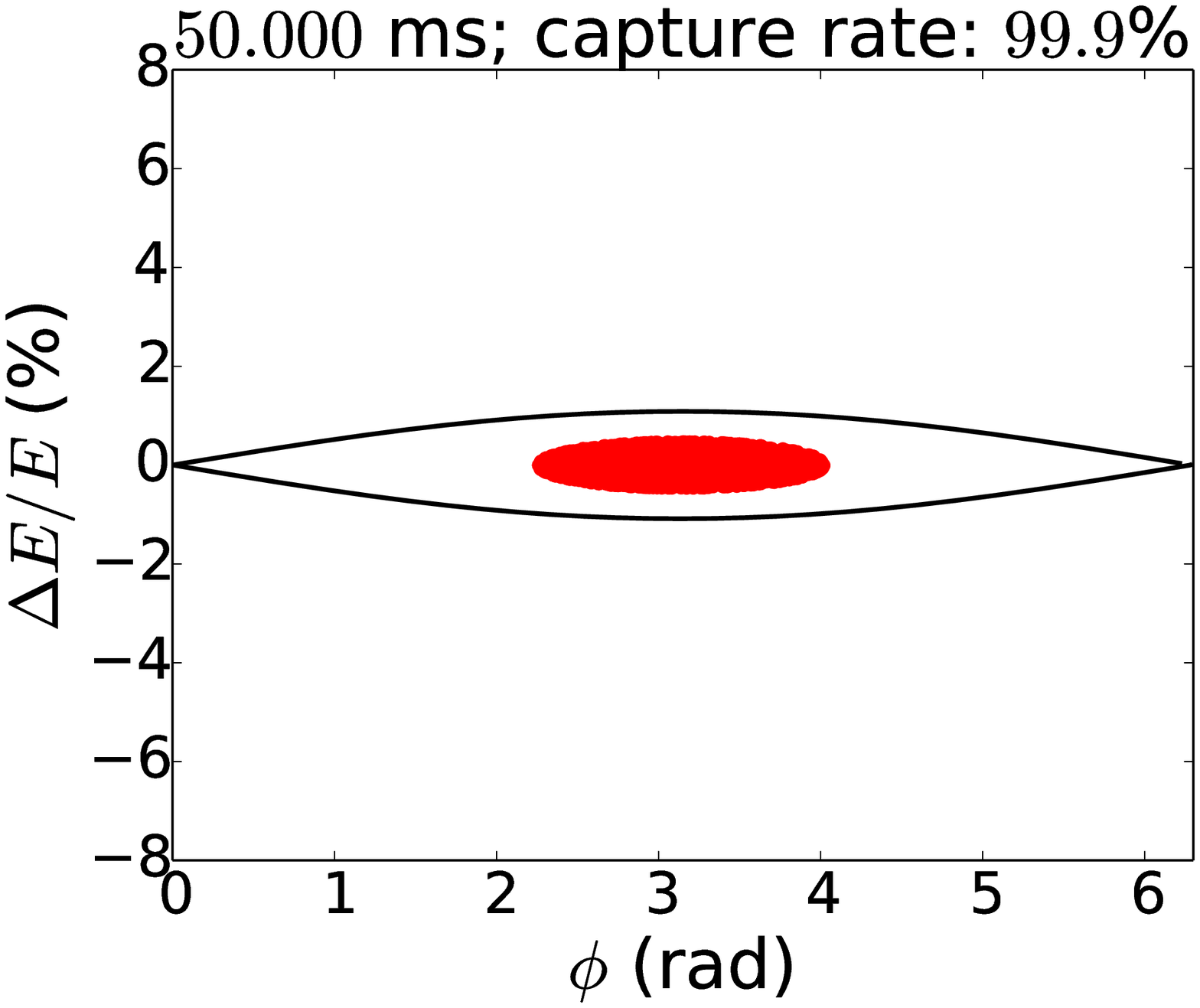,width=40mm,height=40mm}    
    \epsfig{file=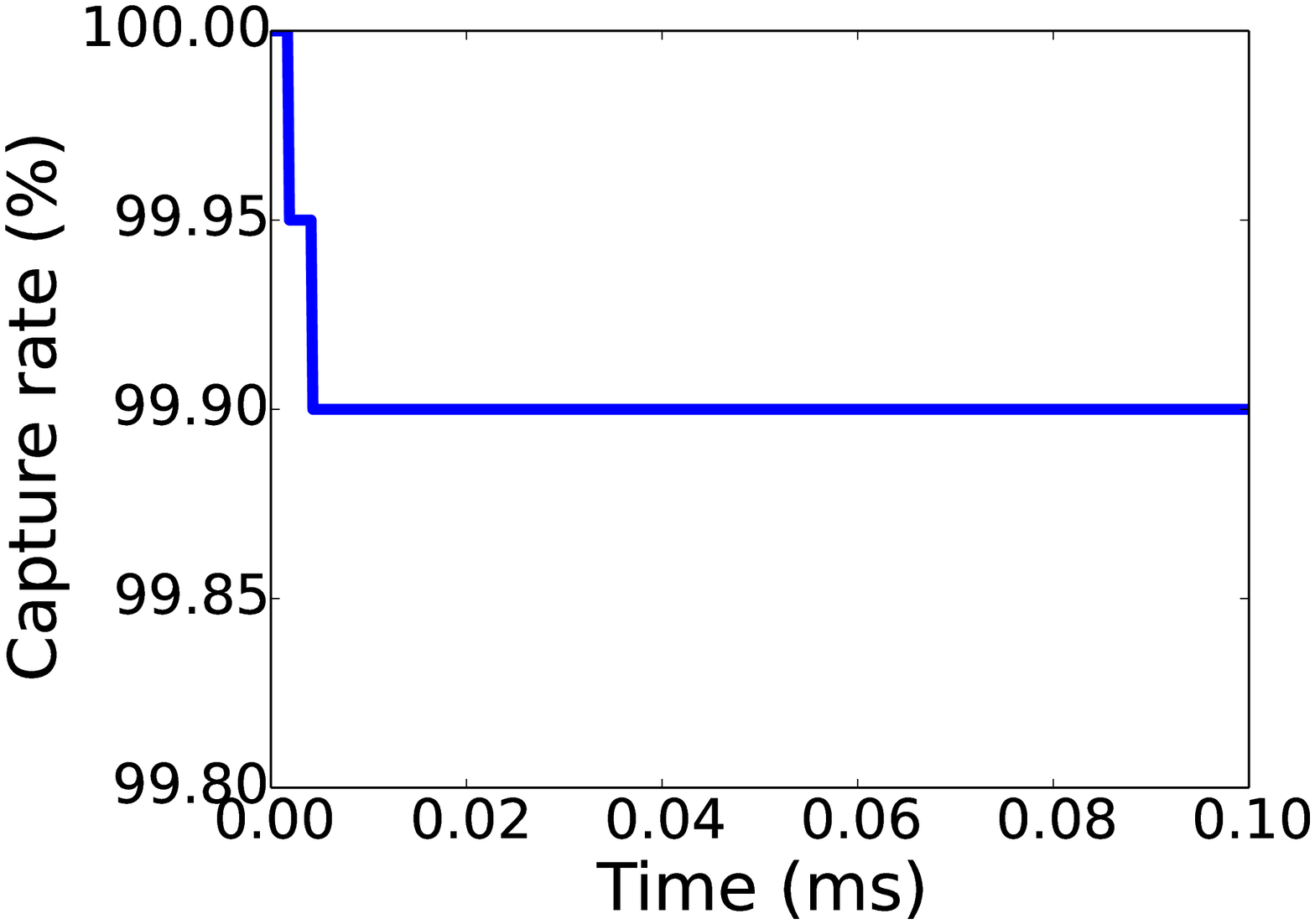,width=60mm,height=45mm}
    \caption{The evolution of phase-space $(\phi,\Delta E/E)$ for 
             electrons in a ramping cycle. 
             We set $T_{\nu}=0.1$ (ms), $V_i=12.5$ (kV) and 
             assume the initial conditions for a bunch of the beam is 
             flatly distributed in the phase $\phi=[0,~2\pi]$ (rad), 
             and the $\Delta E/E$ is a Gaussian distribution with zero mean 
             and $\sigma=\pm0.05\%$, i.e., the width of $\Delta E/E$ is 0.1\%.
             Bottom plot is the capture rate vs. time for electrons in a ramping cycle.}
    \label{fig-electron-phase}
\end{figure}

\begin{figure}[htb]
    \centering
    \epsfig{file=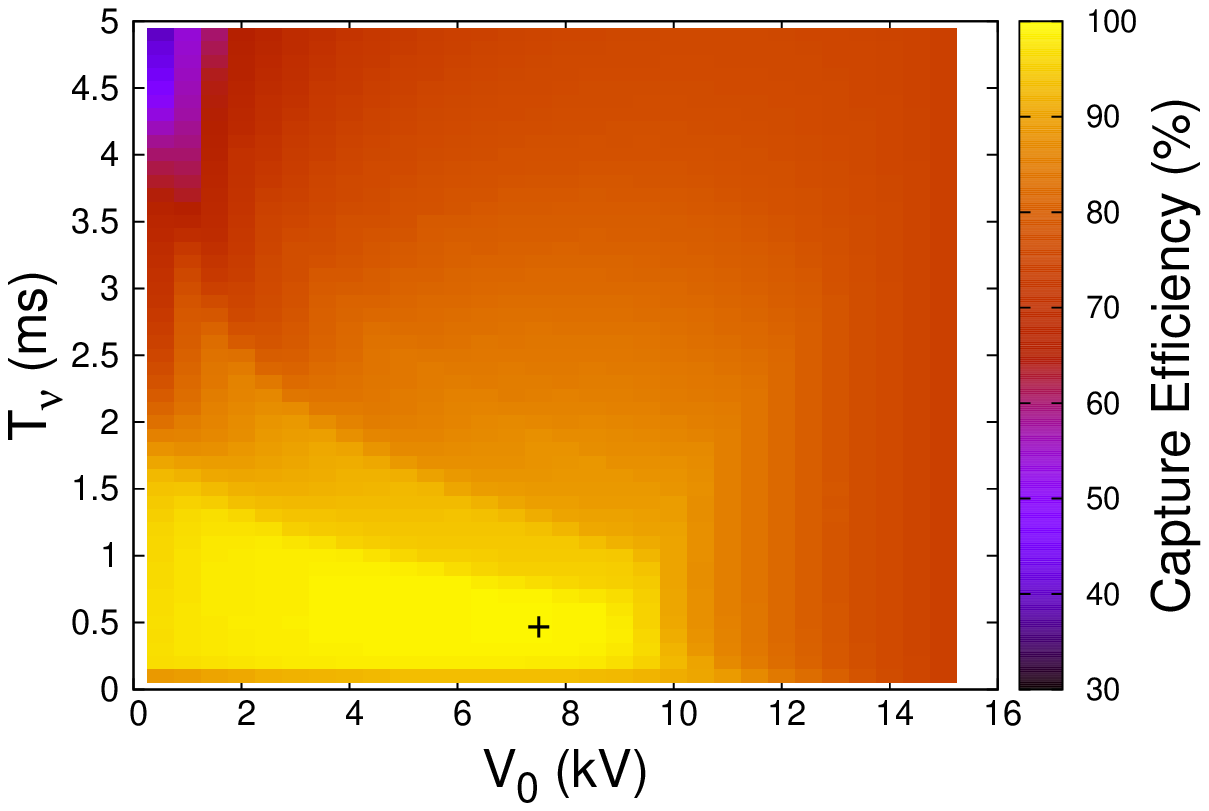,width=80mm,height=60mm}
    \epsfig{file=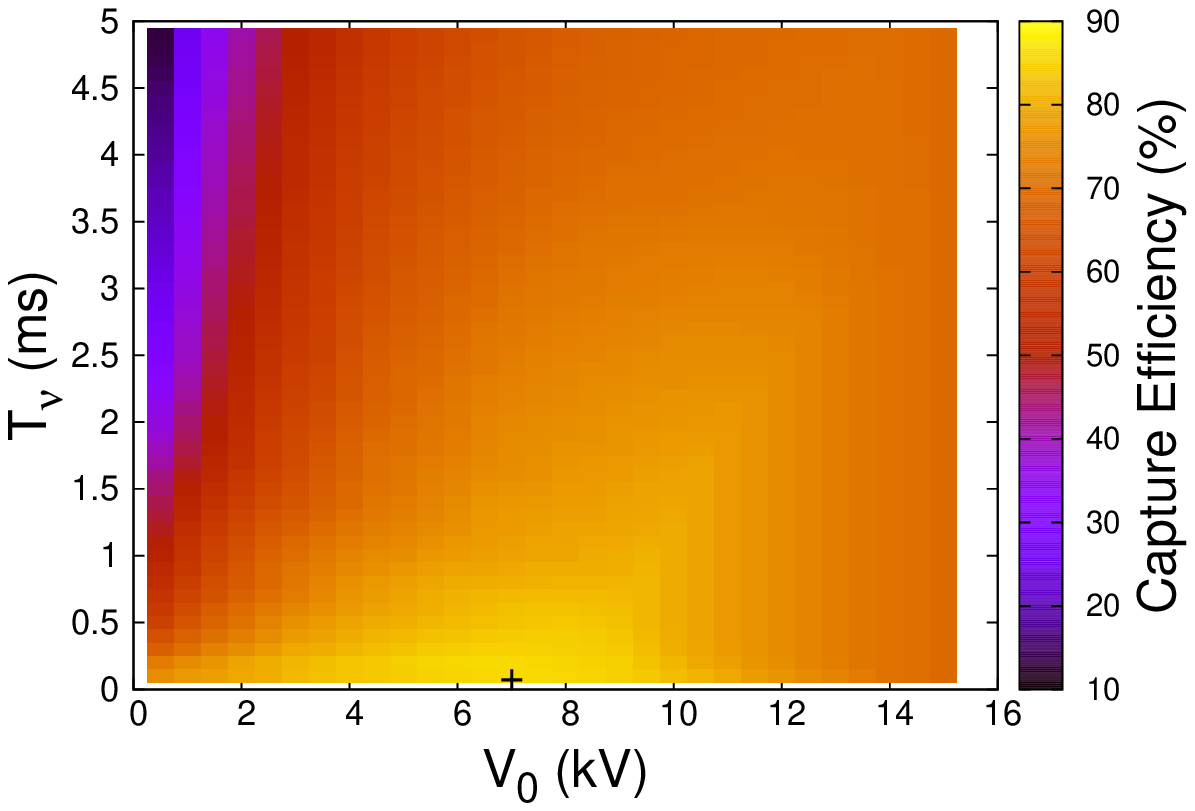,width=80mm,height=60mm}
    \caption{The adiabatic capture efficiencies with respect to 
             RF voltage settings $T_{\nu}$ and $V_i$ for proton beam with Gaussian widths 
             $\Delta P/P=0.1\%$ (upper) or $\Delta P/P=1\%$ (lower).
             The cross marker indicates the position of greatest efficiency 
             in the heat map.}
    \label{fig-proton-eff}
\end{figure}

\begin{figure}[htb]
    \centering
    \epsfig{file=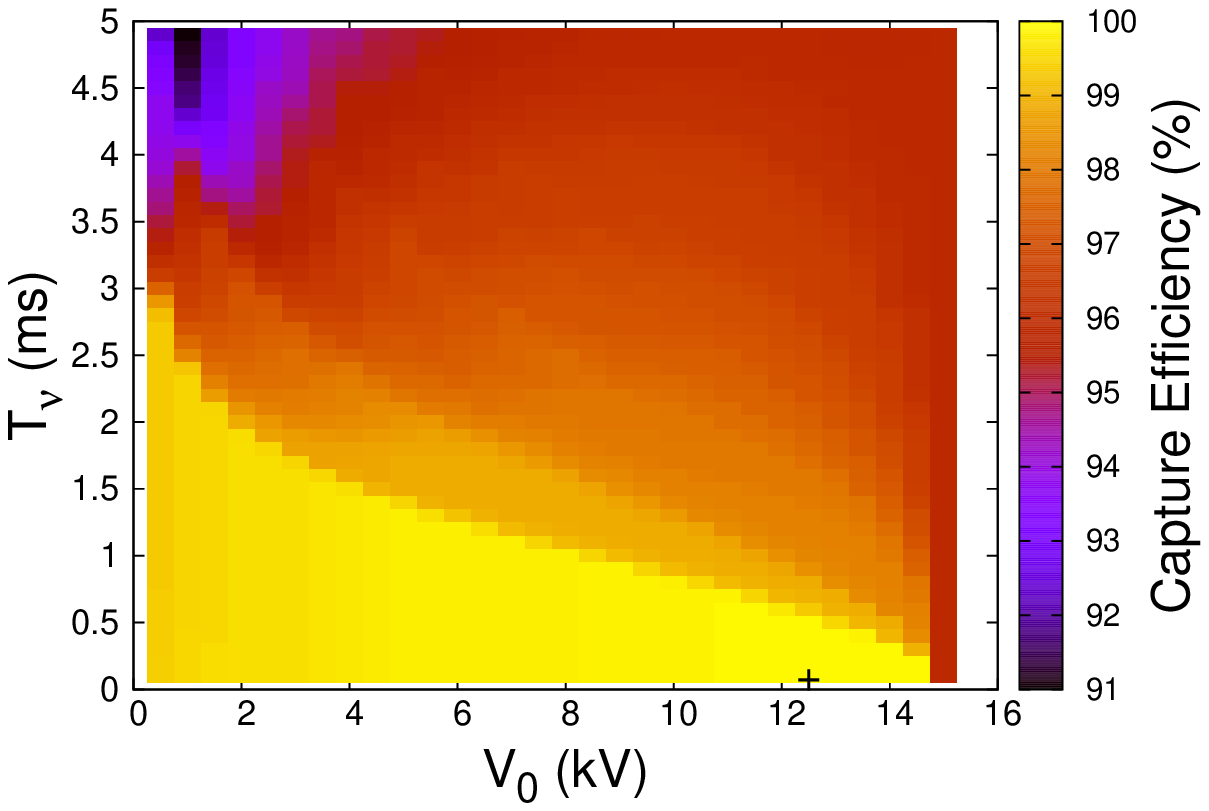,width=80mm,height=60mm}
    \epsfig{file=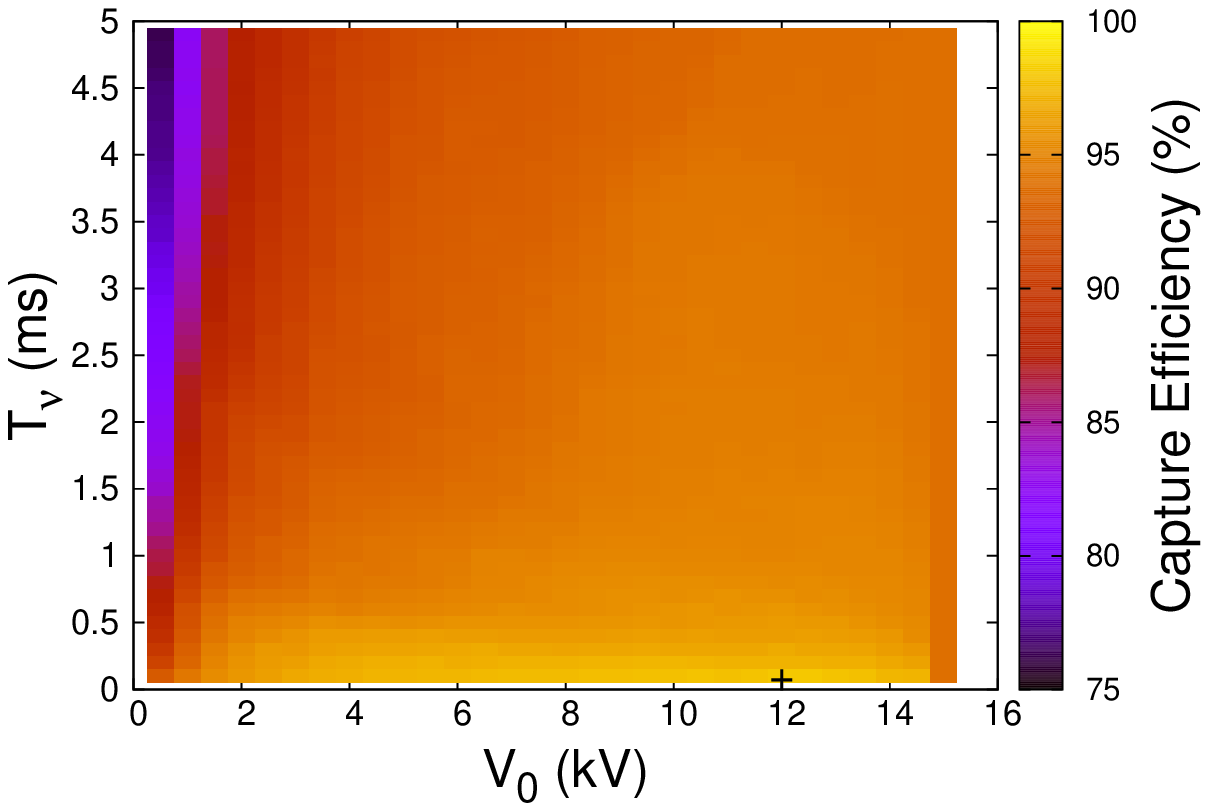,width=80mm,height=60mm}
    \caption{The adiabatic capture efficiencies with respect to 
             RF voltage settings $T_{\nu}$ and $V_i$ for electron beam with Gaussian widths 
             $\Delta E/E=0.1\%$ (upper) or $\Delta E/E=1\%$ (lower).
             The cross marker indicates the position of greatest efficiency 
             in the heat map.}
    \label{fig-electron-eff}
\end{figure}

For protons (electrons) acceleration, 
it takes about 95,500 (208,151) turns to accomplish a ramping cycle.
The properties of adiabatic capture for TLS booster 
as a proton (electron) machine are shown 
in the Fig.~\ref{fig-proton-ramp} (Fig.~\ref{fig-electron-ramp}).
We scan the ranges of $T_{\nu}=[0.1,~4.9]$ (ms) and $V_i=[0.5,~15]$ (kV), 
obtain the capture efficiencies and find the best settings of 
$T_{\nu}$ and $V_i$ for protons (electrons) in the TLS booster.
Fig.~\ref{fig-proton-phase} (Fig.~\ref{fig-electron-phase})
shows an example of the evolution of phase-space and capture rate 
for protons (electrons) in a ramping cycle.
Fig.~\ref{fig-proton-eff} (Fig.~\ref{fig-electron-eff})
shows the heat map of adiabatic capture efficiencies 
for protons (electrons) with respect to $T_{\nu}$ and $V_i$ settings.
Table~\ref{table-TLS-eff} lists the RF voltage settings with $T_{\nu}$ and $V_i$ 
for the best efficiencies of TLS booster as a proton or electron accelerator, 
and with different initial beam distributions $\Delta P/P$ or $\Delta E/E$.

\begin{table}[h]
\begin{center}
\caption{The best capture efficiency with RF voltage settings 
$T_{\nu}$ and $V_i$ for TLS booster}
\begin{tabular}{lccc} 
\toprule
\textbf{Beam type} & \textbf{$T_{\nu}$} & \textbf{$V_i$} & \textbf{Efficiency} \cr
\textbf{(initial Gaussian width)} & \textbf{(ms)} & \textbf{(kV)} & \textbf{(\%)} \cr
\toprule
\small protons ($\Delta P/P=0.1\%$)   & 0.5 & 7.5  & $99 \pm 2.2^*$ \cr
\small protons ($\Delta P/P=1\%$)     & 0.1 & 7    & $85.7 \pm 2.2^*$ \cr
\toprule
\small electrons ($\Delta E/E=0.1\%$) & 0.1 & 12.5 & $99.9 \pm 2.2^*$ \cr
\small electrons ($\Delta E/E=1\%$)   & 0.1 & 12   & $98 \pm 2.2^*$ \cr
\toprule
\multicolumn{4}{l}
{\small * The standard error = $\sqrt{N}/N$, where $N=2000$ is the total} \cr
\multicolumn{4}{l}
{\small \ \ \ number of particles in simulation.} \cr
\end{tabular}
\label{table-TLS-eff}
\end{center}
\end{table} 


\section{SUMMARY}
We have taken the TLS booster as an example to study the adiabatic capture 
efficiency for beam energy ramping.
For protons acceleration, it is better to have the initial beam condition 
with a small $\Delta P/P$ distribution and a small adiabatic capture time 
$T_{\nu}$ for RF voltage setting, in order to have the best efficiency.
For electrons acceleration, the requirements for good capture efficiency 
is roughly same with protons. 
However the electrons acceleration usually has a better efficiency 
compared to protons or other heavy particles, 
as shown in Table~\ref{table-TLS-eff}.

The synchrotron motion simulator is written in Python language 
which is easy to understand and execute on many OS platforms.
It can be freely download from~\cite{ref-CCChiang}. 
This program could be further developed for other kinds of particle 
accelerators, like carbon or heavy ion accelerators, 
for their design and optimization.


\end{document}